\documentclass[prd,preprintnumbers,floatfix,aps,nofootinbib,notitlepage,showpacs]{revtex4-1}

\usepackage{latexsym}
\usepackage{epsfig}
\usepackage{amssymb}

\newcommand{\lp}{\left(}
\newcommand{\rp}{\right)}
\newcommand{\lb}{\left[}
\newcommand{\rb}{\right]}

\newcommand{\ba}{\begin{eqnarray}}
\newcommand{\ea}{\end{eqnarray}}
\newcommand{\be}{\begin{equation}}
\newcommand{\ee}{\end{equation}}
\newcommand{\bk}{{\bf k}}

\newcommand{\al}{\alpha}
\newcommand{\bt}{\beta}
\newcommand{\ga}{\gamma}

\newcommand{\la}{\lambda}

\newcommand{\en}{\epsilon}

\newcommand{\Ga}{\Gamma}
\newcommand{\Sa}{\Sigma}

\newcommand{\GG}{\gamma^2}

\newcommand{\h}{\mathcal{H}}

\begin{document}

\title{Quantum backreaction in evolving FLRW spacetimes}
\date{\today}

\author{Tomi S. Koivisto}
\email{T.S.Koivisto@uu.nl}
\author{Tomislav Prokopec}
\email{T.Prokopec@uu.nl}
\affiliation{Institute for Theoretical Physics and Spinoza Institute, Leuvenlaan 4, 3584 CE Utrecht, The Netherlands.}

\preprint{ITP-UU-10/34}
\preprint{SPIN-10/29}


\begin{abstract}

Quantum fluctuations of a nonminimally coupled scalar field in D-dimensional homogeneous and isotropic background are
calculated within the operator formalism in curved models with time evolutions of the scale factor that allow smooth transitions between contracting and expanding and between decelerating and accelerating regimes.
The coincident propagator is derived and used to compute the one-loop backreaction from the
scalar field.
The inflationary infrared divergences are absent in Bunch-Davies vacuum when taking into account a preceding
cosmological era or
spatial curvature which can be either positive or negative. It is found that asymptotically, the backreaction energy
density in the minimally coupled case grows logarithmically with the scale factor in quasi-de Sitter space, and in a class of models
decays in slow-roll inflation and grows as a power-law during super-inflation.
The backreaction increases generically in a contracting phase or in the presence of a negative nonminimal
coupling. The effects of the coupling and renormalization scale upon the quantum fluctuations together with the
novel features due to nontrivial time evolution and spatial curvature are clarified with exact solutions and numerical
examples.

\end{abstract}

\maketitle


\section{Introduction}

At the largest scales, the universe seems very nearly homogeneous and isotropic. Thus its evolution can be described by a single scale factor $a(\eta)$, whose time evolution is parameterized by the conformal time $\eta$. The metric, originally studied by Friedmann, Lema\^itre, Robertson and Walker (FLRW), can be written as
\be \label{metric}
g_{\mu\nu}dx^\mu dx^\nu = a^2(\eta)\lp -d\eta^2 +  \frac{dr^2}{1-Kr^2} + r^2d\theta^2 + r^2\sin^2\theta d\phi^2\rp\,.
\ee
The curvature of the spatial hypersurfaces is given by the constant $K$. The current cosmological measurements \cite{Komatsu:2008hk} give a very large lower limit on the radius $R_c=|K|^{-\frac{1}{2}}$ of the curvature of the universe, $R_c \ge 22h^{-1}$ Gpc ($R_c \ge 33h^{-1}$ Gpc) for a universe with positively (negatively) curved spatial sections, where $h=0.705 \pm 0.013$ is the Hubble parameter evaluated today in units of 100 km/s/Mpc. In the present work we will work in terms of the conformal Hubble rate defined as
\be
\h(\eta) \equiv \frac{d}{d\eta}\log{\lp a(\eta)\rp}\,.
\ee
The leading paradigm explaining the observed approximate isotropy, homogeneity and flatness of the universe is inflation \cite{Guth:1980zm}. Remarkably, an inflationary theory provides also an origin for the observed structures in the universe as a result of quantum fluctuations in expanding space \cite{Starobinsky:1982ee}.
\newline
\newline
Scalar fields have been ubiquitous in the studies of this phenomenon. The behavior of a massless scalar field
with a possible non-minimal coupling to the Ricci scalar on expanding spaces \cite{Vilenkin:1982wt} has been
studied extensively. This is relevant also for understanding of higher spin fields. In particular the graviton can be related to the kinetic operator of a scalar field. Apart from some tensorial structure, the graviton propagator can be written in terms of propagators of massless scalar fields \cite{Grishchuk:1974ny} also in quasi-de Sitter space \cite{Janssen:2008dw}, and similar observations can be made for vector fields \cite{Tsamis:2006gj} or antisymmetric tensor fields \cite{Janssen:2007yu}. It is also shown that a three-form inflation in some limits allows an equivalent scalar field description \cite{Germani:2009gg,Koivisto:2009fb}. Therefore the understanding of the massless scalar field can be translated to more general fields as well.
\newline
\newline
The key property of quantum behavior of fields in an expanding spacetime is the particle production that stems from the presence of a horizon. The particles are created mostly in the infrared, and it is this effect that leads for example to the creation of primordial density fluctuations. This also raises the issue of the backreaction of the fluctuations on the background spacetime \cite{Tsamis:1996qm,Mukhanov:1996ak}. When one chooses the vacuum to contain only purely positive frequency modes (the so called Bunch-Davies vacuum \cite{Bunch:1977sq}), the expectation value of the two point correlator for this vacuum diverges in the infrared. Unlike the ultraviolet divergences, which can be removed by standard procedures, these infrared divergences show that the system under study is pathological. Hence, it is clear that the Bunch-Davies vacuum\footnote{Here our definition of the Bunch-Davies vacuum differs from the one used in e.g. Ref. \cite{bd}, where the positive frequency solutions are defined at the limit $\eta\rightarrow-\infty$. For calculational purposes it is impractical to take this limit when the wavenumbers $k$ considered range from $0$ to $\infty$, and we pick the positive frequency modes at a finite $\eta_0$.} cannot describe a physically sensible state of an exponentially expanding spacetime. Possibilities that emerge are to consider an alternative vacuum state with less singular superhorizon modes \cite{Vilenkin:1983xp} or a compact spatial manifold where those modes are initially absent \cite{Tsamis:1993ub}.
\newline
\newline
Many studies though concern the special limit of exactly exponential inflation, i.e. de Sitter background \cite{Linde:1982zj,Losic:2006ht,Losic:2008ht}. It is natural to consider more general evolutions to understand how the predictions from more realistic assumptions may differ from the maximally symmetric case. Recently the scalar and graviton propagators have been derived for flat FLRW universe with constant deceleration parameter \cite{Janssen:2007ht}. The infrared divergencies persist for a large and physically relevant range of the deceleration parameter, in particular for accelerating regime if the field is minimally coupled. However, in Ref. \cite{Janssen:2009nz} it was shown that the infrared can be regulated by taking into account a decelerating epoch preceding inflation. The perturbation modes were matched at a transition from radiation domination to inflation, and the results agreed qualitatively with the ones obtained by using a cut-off procedure in Ref. \cite{Janssen:2008px}, when the effects of the particle creation due to the sudden transition approximation were properly identified.
\newline
\newline
In this paper we consider universes with a smooth transition to an inflationary period. This is achieved by considering more general background evolution specified by three parameters, which still allow analytic expressions for the solutions of the mode functions. Furthermore, we also include the possibility of nonzero spatial curvature. It is natural to consider that the universe was not exactly flat but the curvature radius was inflated beyond the present observational limits. In such a case, we cannot observe the curvature directly but we obtain consistent results from the calculation of fluctuations. In a similar way, by taking into account a bouncing or a decelerating phase preceding inflation will rather generically eliminate the infrared infinities in the quantum fluctuations. Thus it turns out that in the more complete models featuring either some curvature or realistic time-evolution, the inflationary spectra are regular. As some details of the spectra depend on the physical means of regularization, there is also a hope to obtain some information on pre-inflationary cosmology.
\newline
\newline
Having then well-defined and regular quantum fluctuations allows to investigate their physical effects. It would
be of interest to extend the study to the graviton. In this work we however focus on the leading order
corrections from the massless scalar field in particular during inflation. For this purpose we derive the
one-loop expectation
value of the induced stress energy tensor for a scalar field. Our results confirm the well-known leading
logarithm behavior in quasi-de Sitter inflation \cite{Vilenkin:1982wt,Starobinsky:1982ee,Linde:1982uu} (see the
interesting recent discussions on the physical relevance of such logarithms and their implementation within the
$\delta N$ formalism \cite{Giddings:2010nc,Byrnes:2010yc}, also \cite{Senatore:2009cf,Kahya:2010xh} and the
review \cite{Seery:2010kh}.) If the backreaction grows, the perturbative treatment will eventually become
invalid. There are suggestions that such may signal analogous breakdown of locality in quantum gravity in
cosmological as in black hole physics \cite{ArkaniHamed:2007ky}. We find that the backreaction generically
increases also when the scalar field has a negative coupling to the Ricci curvature, in agreement with the conclusion reached in Ref. \cite{Janssen:2009nz}, or when the universe is contracting or super-accelerating. In most cases the sign of the induced energy density can change dynamically and depends on the parameters of the model.
\newline
\newline
We set up our notation with some preliminary considerations in section \ref{scalar_sec} where we also describe the
background. The propagator is derived in section \ref{back_sec} using the operator formalism. For the above mentioned
reasons, particular care is taken to obtain physically regular infrared, since the divergences (except
logarithmic ones) would be absorbed by dimensional regularization. The ultraviolet is renormalized by counterterms. In section \ref{ba_sec} we then apply the propagator to compute the backreaction from quantum fluctuations. In particular, the asymptotically dominating contribution at inflation is extracted. Finally, we briefly discuss the conclusions in section \ref{conc_sec}. Some details of the $D$-dimensional curved FLRW models are confined to the appendices: in appendix \ref{field_app} we construct the curvature from the metric and derive the field equations, in appendix \ref{k_app} we state the spherical Laplacian operator and discuss its eigenmodes. Appendix \ref{special} contains explicit expressions for the propagator in a few special cases.

\section{Scalar field in expanding space}
\label{scalar_sec}

 Here, in \ref{field} we will sketch the usual quantization of a scalar field in FLRW universe, and propose the ansatz (\ref{ansatz}) for its background evolution. We then describe this class of background evolutions in \ref{background}.

\subsection{Scalar field}
\label{field}

Consider a massless non-minimally coupled scalar field $\Phi$ in a $D$ dimensions,
\be \label{action}
S= \frac{1}{2}\int d^Dx\sqrt{-g}\Phi(\Box-\xi R)\Phi\,.
\ee
Here $R$ is the Ricci scalar and $g$ the determinant of the metric. The d'Alembertian operator is denoted as $\Box = \nabla^\mu\nabla_\mu$. The field obeys a simple equation of motion
\be \label{eom1}
\lp \Box-\xi R \rp\Phi = 0\,.
\ee
The stress energy tensor becomes
\be \label{emt}
T_{\mu\nu} \equiv \frac{-2}{\sqrt{-g}}\frac{\delta S}{\delta g^{\mu\nu}} = \Phi_{,\mu}\Phi_{,\nu}-\frac{1}{2}g_{\mu\nu}(\partial\Phi)^2 + \xi\left(G_{\mu\nu}-\nabla_\mu\nabla_\nu+g_{\mu\nu}\Box\right)\Phi^2\,,
\ee
with the trace
\be \label{trace}
T \equiv g^{\mu\nu}T_{\mu\nu} =-\frac{D-2}{2}\left((\partial\Phi)^2+\xi R\Phi^2\right)+(D-1)\xi\Box\Phi^2\,.
\ee
We shall be interested in deriving the Feynman propagator $i\Delta(x;x')$, which satisfies the equation
\be \label{eom}
\sqrt{-g}\lp \Box -\xi R\rp i\Delta(x;x') = i\delta(x-x')\,.
\ee
For this purpose we work in the operator formalism, and eventually sum over the mode functions (solutions of the equation (\ref{eom1}) in Fourier space) to compute the propagator. This way we obtain the complete answer, whereas a solution to the Eq. (\ref{eom}) is unique only up to the homogeneous modes.
\newline
\newline
The metric, regarded as a background, was introduced in Eq. (\ref{metric}). Due to isotropy, the Fourier components of the field can then be expanded with the creation and annihilation operators $b_{\bk}$ and $b^\dagger_\bk$ as
\be \label{phi}
\Phi_\bk(\eta) = \psi_k(\eta)b_\bk + \psi^*_k(\eta)b_{-\bk}^\dagger\,.
\ee
The convention we use for the transformation is
\be
F({\bf x}) = \frac{1}{(2\pi)^{\frac{D-1}{2}}}\int d^{D-1}k F_{\bf k}Q({\bf k},{\bf x})\,.
\ee
In a closed universe the integral is more properly replaced by a sum. In flat universe the mode functions are simply plane waves, $Q({\bf k},{\bf x})=e^{i{\bf k}\cdot{\bf x}}$, and some details of the curved cases are given in appendix \ref{k_app}. So the comoving wavelengths of perturbations are $\sim a/k$.
The canonical commutation relations for the operators and the field,
\ba
\left[ b_{\bk},b^\dagger_{\tilde{\bk}} \right] & = & (2\pi)^{D-1}\delta^{D-1}(\bk-\tilde{\bk})\,, \\
\left[\Phi_\bk(\eta),a(\eta)^{D-2}(\Phi^{\dagger}_{\tilde{\bk}})'(\eta) \right] & = & i(2\pi)^{D-1}\delta^{D-1}(\bk-\tilde{\bk})\,,
\ea
imply that the Wronskian of the mode functions is given by
\be \label{wronskian}
\psi_k(\eta)(\psi^{*}_k)'(\eta)- \psi^*_k(\eta)\psi_k'(\eta) = ia^{2-D}(\eta)\,.
\ee
Above and in the following, a prime indicates a derivative with respect to the conformal time, $' = d/d\eta$. From now on we will occasionally drop the subscripts $k$ and arguments of $\eta$.
\newline
\newline
To present Eq. (\ref{eom1}) in the metric (\ref{metric}), we need the Ricci scalar (\ref{ricciscalar}) from appendix \ref{field_app} and the d'Alembertian operator (\ref{dalembertian}) from appendix \ref{k_app}.
Notably, the Laplacian part of the latter, when acting upon the Fourier modes, gives a combination reducing to $(k^2-K)$ in $D=4$.
By the end of the day, the mode functions $\psi_k$ are found to satisfy the evolution equation
\be \label{p_evol}
\lb \partial_\eta^2+k^2+\lp -\frac{1}{4}(D-2) + (D-1)\xi\rp(D-2) K + M^2(\h)\rb\lp \psi_k a^{\frac{D}{2}-1}\rp=0\,,
\ee
where we have lumped the Hubble-dependent terms into the definition
\be
M^2(\h) = -\frac{1}{2}\lp D-2 \rp \lp \h'+ \frac{1}{2}\lp D-2\rp \h^2 \rp + (D-1)\lp 2\h'+\lp D-2\rp\h^2\rp\xi \,.
\ee
Because this allows us an analytic solution for the mode functions, we assume that
\be \label{ansatz}
M^2(\h) = \lp\GG + \frac{\frac{1}{4}-\nu^2}{\eta^2}\rp\,,
\ee
where $\GG$ and $\nu$ are $D$-independent constants. This parameterization captures a wide range of interesting
geometries in
$D=4$ as described in the following subsection. The generalization of the geometries to higher dimensions is by no means unique. One could add arbitrary terms proportional to $(D-4)$ into equation (\ref{ansatz}), but we chose the trivial $D$-dependence for simplicity. Such choices in an analytic extension affect only the finite constants that appear in the results. Note that after we have dictated (\ref{ansatz}), the rest of the terms in the evolution equation (\ref{p_evol}) are uniquely determined by the action (\ref{action}). The remaining ambiguity will appear in the choice of the $D$-dependence of the initial conditions for the solutions to this equation. This ambiguity is fixed by the natural prescription we adopt in section \ref{pr_sec}, namely that in any $D$ the modes are initially in their Bunch-Davies state.

\subsection{Background}
\label{background}
\begin{figure}[ht]
\includegraphics[width=8.5cm]{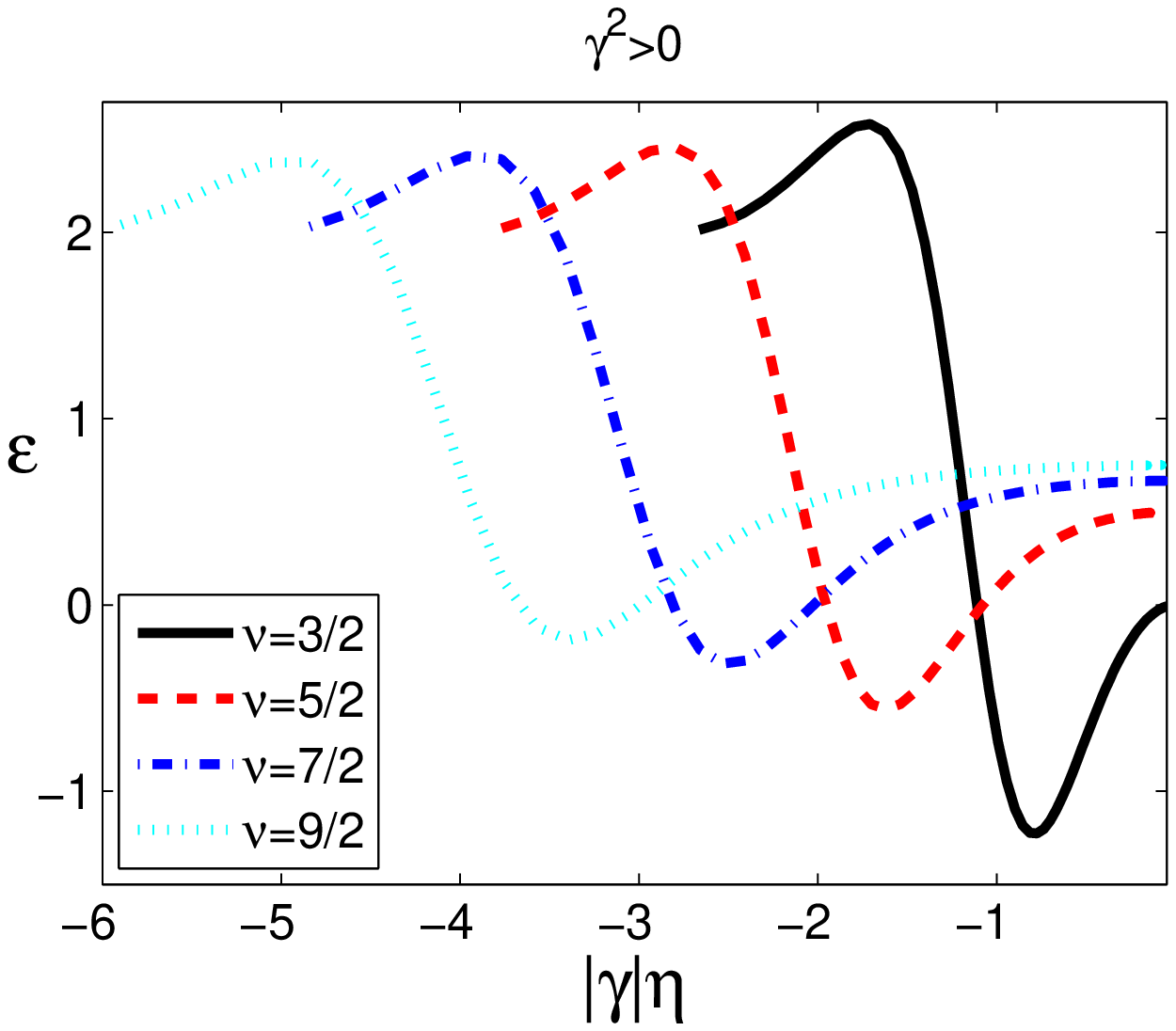}
\includegraphics[width=8.5cm]{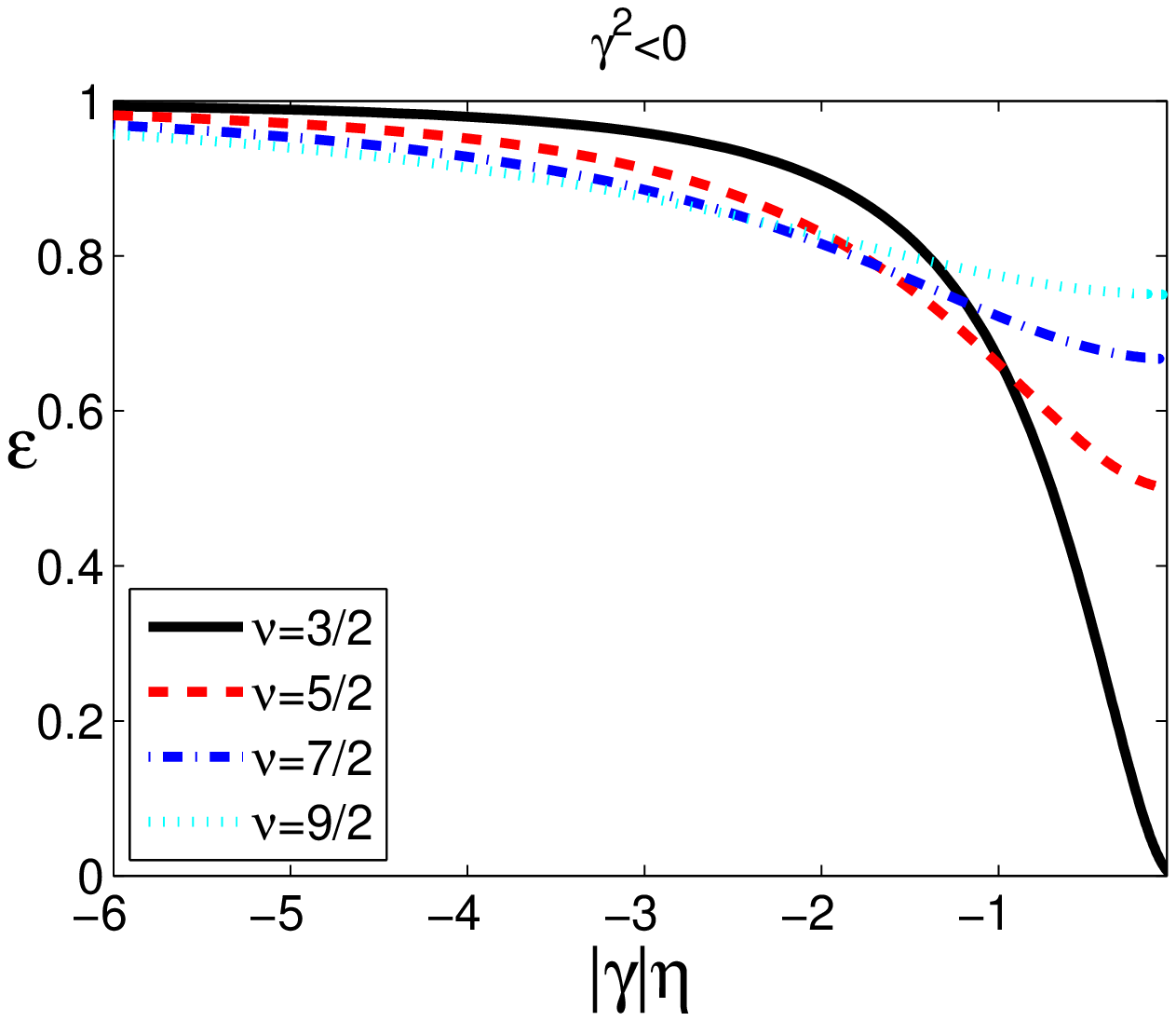}
\caption{\label{eps1}
Evolution of the slow-roll parameter $\epsilon$ as a function of the conformal time $\eta$ in units of $|\gamma|$.
{\bf Left}: Solutions to Eq. (\ref{a_evol}) with positive $\ga^2$. The curves start from the beginning of the universe.
{\bf Right}: Solutions to Eq. (\ref{a_evol}) with negative $\ga^2$. The curvature dominated phase can be extrapolated to $\eta\rightarrow -\infty$.}
\end{figure}
Let us look at which kinds of background evolution the ansatz (\ref{ansatz}) allows. In $D=4$, the evolution equation for the scale factor becomes
\be \label{a_evol}
a'' =  -\frac{a}{1-6\xi}\lp  \GG + \frac{\frac{1}{4}-\nu^2}{\eta^2}\rp\,.
\ee
We can set $\xi=0$ for simplicity. A nonminimal coupling corresponds then to redefining the parameters as
\begin{equation} \label{xi_scale}
\nu \rightarrow \nu_\xi  =  \sqrt{\frac{\nu^2-\frac{3}{2}\xi}{1-6\xi}}\,, \quad
\gamma \rightarrow \gamma_\xi  =  \frac{\gamma}{\sqrt{1-6\xi}}\,,
\end{equation}
when $\xi<\min(\frac{1}{6},2\nu^2/3)$.
One sees that though the asymptotic behaviors at large and small $\eta$ are the same as in a cosmology with curvature and an energy component with the constant equation of state $w=(3\pm 2\nu)/(3\mp 6\nu)$, the geometry of (\ref{ansatz}) is not precisely reproduced by the latter configuration but generally requires a dynamical $w$. In the constant-$w$ case, Eq. (\ref{ansatz}) would be replaced by
\be
\frac{a''}{a}=K+\frac{3}{2}(1-3w)|K|S^{-2}\left[\frac{1}{2}(1+3w)\sqrt{|K|}\eta\right]\,,
\ee
where $S(x)=\sinh{x}$ for negative and $S(x)=\sin{x}$ for positive curvature. Since $S(x) \approx x + \dots$, our results should qualitatively describe also the
particular case of spatial curvature and a constant-$w$ fluid as the energy sources. In more detail, the evolutions we have that lead to inflation can be classified into three qualitatively different types as we describe below. We will later need the scalar curvature for these geometries, which is
\be \label{curvature}
R = -\frac{6}{(1-6\xi)a^2}\left[\gamma^2+\frac{\frac{1}{4}-\nu^2}{\eta^2}-(1-6\xi)K\right]\,,
\ee
when $D=4$.

\subsubsection{$\ga^2>0$: From radiation domination to inflation}
\label{pos}

In the explicit examples here, we assume that $\nu$ is a half-integer $\nu>\frac{1}{2}$ but this assumption is not crucial to our conclusions as explained later. Let us first consider the case of real $\ga>0$. Then the exact solution to (\ref{a_evol}) can be expressed in terms of the Bessel functions
\be \label{bes}
a(\eta) = (-1)^{\nu-\frac{1}{2}}\frac{\sqrt{\ga\eta}\pi}{2}\lb c_1 J_{\nu}(\gamma \eta) + c_2  Y_{\nu}(\gamma \eta )\rb \,,
\ee
where $c_1$ and $c_2$ are real constants.
In an accelerating universe, $\eta$ approaches zero from below. At late times, from the asymptotics of the Bessel function, we see that $a(\eta) \rightarrow |c_2|
2^{\nu-1}\Gamma(\nu)|\ga\eta|^{\frac{1}{2}-\nu}$, so the solutions describe the evolution infinitely far in the future
where the scale factor grows without limit when $\nu>\frac{1}{2}$. This power-law evolution implies that the deceleration parameter will be asymptotically a constant, which is given by $\nu$ as $\epsilon = (3\pm 2\nu)/(1\pm 2\nu)$. In this paper we consider the upper sign choice since we are interested in inflation, but it should be kept in mind that for example $\nu=3/2$ can correspond to matter dominated era as well as de Sitter spacetime. We use
\be
\epsilon \equiv 1-\frac{\h'}{\h^2}\,,
\ee
corresponding to the usual definition of the slow-roll parameter. It is sometimes convenient to refer also to the deceleration parameter $q=\epsilon-1$. If the scale factor decelerates with respect to the cosmic time, $q$ is positive, and in the accelerating case $q<0$. These are what we refer to as accelerating or decelerating in this paper (one notes that acceleration with respect to the conformal time $a''>0$ corresponds to rather $\epsilon<2$ and $q<1$).
\newline
\newline
In the early times, the solution will however be oscillating, as immediately seen from Eq. (\ref{a_evol}) when $\xi<1/6$. This means we cannot extend the evolution infinitely far in the past, but there is a beginning to the universe. Inspecting the behavior of the solutions, we see that generically the evolution in the physical region (positive scale factor) begins with the slow-roll parameter near $\en=2$, corresponding to radiation domination. Numerical examples are plotted in the left panel of figure \ref{eps1}. These correspond to such initial conditions that $c_1=0$ and thus
\be \label{hoo_p}
\h = \frac{1-2\nu}{2\eta} + \ga\frac{Y_{\nu-1}(\ga \eta)}{Y_\nu(\ga\eta)}\,.
\ee
We note that a possible feature is a period of super-inflation (negative slow roll parameter) during the transition from $\en \approx 2$ to the constant-$\en$ phase given by the first term in (\ref{hoo_p}).

\subsubsection{$\ga^2 <0$: From curvature domination to inflation}
\label{neg}

Let us then consider an imaginary $\ga$ in equation (\ref{ansatz}). The early behavior then becomes exponentially growing in $\eta$ instead of oscillating. Then $\en=1$, corresponding to a curvature-dominated universe.
The exact solution can then be written in terms of modified Bessel functions, and we can extend the evolution infinitely far in the past. We have
\be \label{modbes}
a(\eta) = \sqrt{|\ga\eta|}\lb c_1 I_{\nu}(|\gamma\eta|) + c_2  K_{\nu}(|\gamma\eta|)\rb \,,
\ee
where $c_1$ and $c_2$ are again real parameters. We plot some example evolutions corresponding to the initial condition $c_1=0$ in the right panel of figure \ref{eps1}.
This corresponds to the conformal Hubble parameter
\be \label{hoo_m}
\h = \frac{1-2\nu}{2\eta} + |\ga|\frac{K_{\nu-1}(|\ga \eta|)}{K_\nu(|\ga\eta|)}\,.
\ee
We chose the normalization in such a way that future asymptotic expansion is the same as in the previous case. It turns out we must include curvature in the case
of imaginary $\ga$ to consistently regulate the infrared. The required amount is $|K|>|\ga^2|$ if $K<0$ and $K>|\ga^2|/3$ if $K>0$, as will be clarified later.


\begin{figure}[ht]
\includegraphics[width=8.5cm]{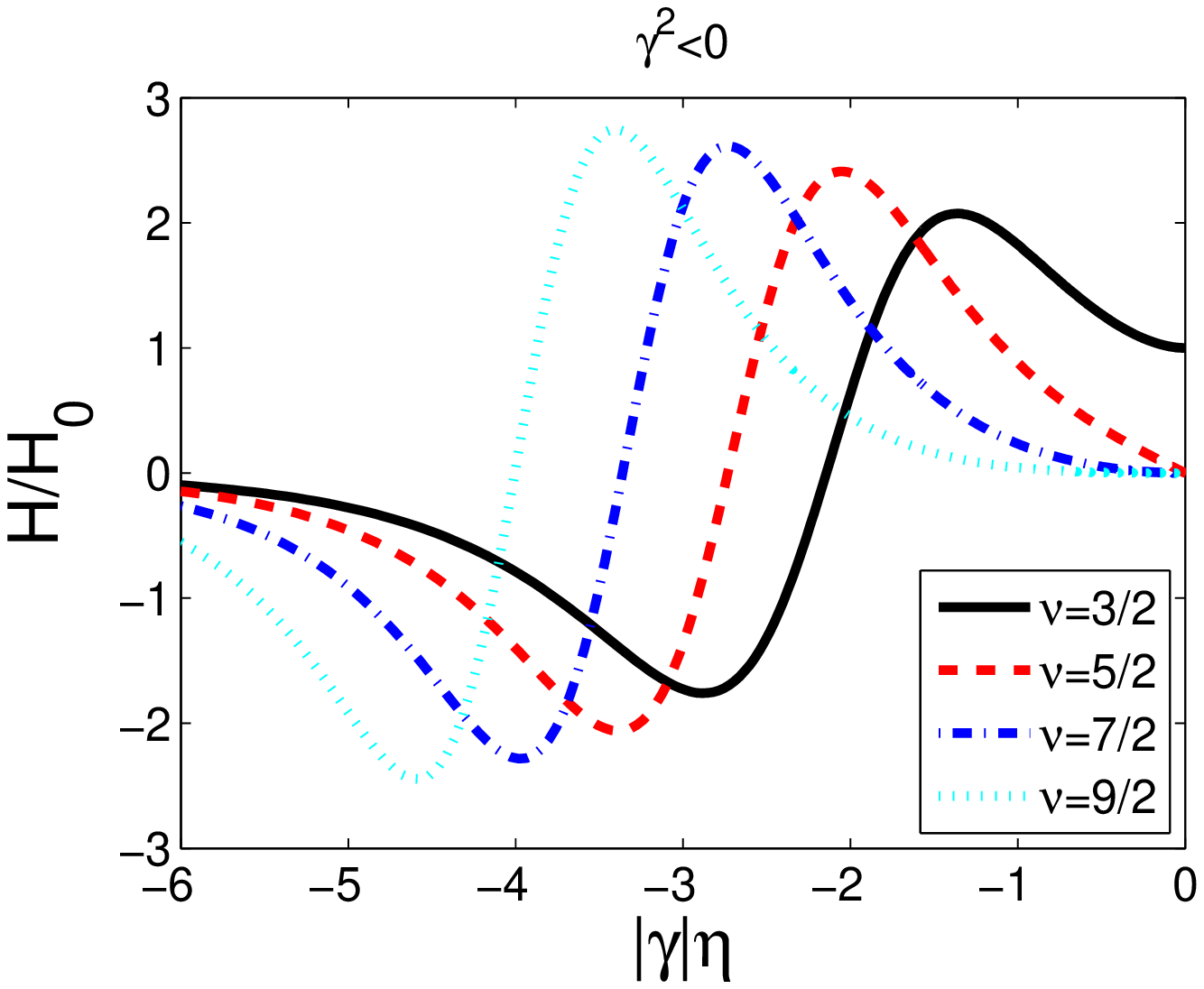}
\includegraphics[width=8.5cm]{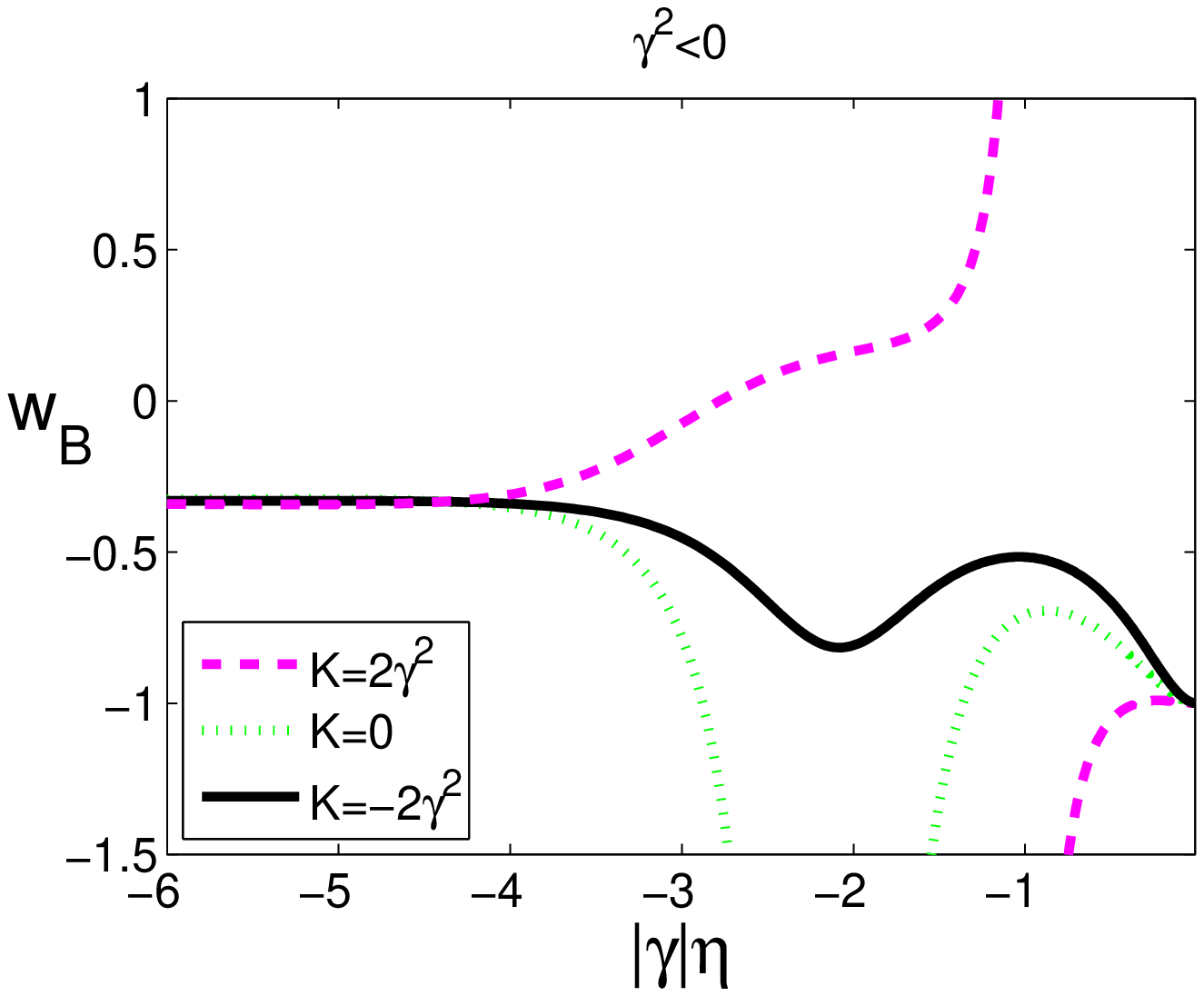}
\caption{\label{eps2}
Bouncing universes described by Eq. (\ref{modbes}) when $\ga^2<0$ and $c_2=10c_1$. Lowering $|\gamma|$ corresponds to stretching the $\eta$-axis.
{\bf Left}: The Hubble rate $H=\h/a$ for several values of $\nu$. The evolution begins from a contracting phase $H<0$, experiences a bounce at $H=0$ and enters an accelerating phase. In case $\nu=3/2$ the latter becomes de Sitter and $H$ approaches a constant $H_0$.
{\bf Right}: The effective equation of state of the background fluid for the $\nu=3/2$ case. In closed universe, the fluid obeys the null energy condition. This is not the case in open or flat universe (and the latter example is not infrared regular).}
\end{figure}
\subsubsection{$c_1,c_2 > 0$: Bouncing from contraction to inflation}
\label{bounce}

A third qualitatively different case arises when we include allow also a nonzero $c_1$ in either the solution (\ref{bes}) or the solution (\ref{modbes}). For the sake of clarity, we consider only the latter case here. Then we find a contracting phase as $\eta\rightarrow -\infty$. Thus the inflationary epoch was preceded by a cosmological bounce. Such scenarios have received lot of attention \cite{Novello:2008ra}, because they may resolve the cosmological singularity and extend the evolution to sort of a pre-Big Bang phase. A bouncing universe can be geodesically complete unlike an eternally expanding one \cite{Borde:2001nh}.
\newline
\newline
Few examples are plotted in figure \ref{eps2}. In the left panel we show the Hubble rate $\h/a$. The right panel depicts the effective equation of state $w_B$ for the energy component that is the source for the Hubble rate. It is well known that in general relativity, only null-energy condition violating energy sources can drive a bounce in a flat universe, whereas in the presence of curvature it can be sufficient to violate the strong energy condition. One may however obtain bounces in flat setting without introducing ghosts in generalized gravity theories; recent progress has been made within the first order formalism \cite{Koivisto:2010jj,Barragan:2010qb} and for string-inspired nonlocal gravity \cite{Calcagni:2010ab,Biswas:2010zk}.
Using the equations in appendix \ref{field_app}, we have deduced the equation of state $w_B$ of the background fluid in the present case.
The general expressions are rather cumbersome, but let us quote the result in the special case $\nu=3/2$:
\be
w_B=
-\frac{c_1^2 \left[t \left(t \left((t+1)^2
   \kappa+2\right)+6\right)+3\right]+2 c_1 \hat{c}
   \lb t^4 (2+\kappa)-t^2 (\kappa-4)-3\rb e^{2 t}  + \hat{c}^2 \lb t \left(t \left((t-1)^2 \kappa+2\right)-6\right)+3\rb e^{4 t} }{3
   c_1^2 \lb t (t+1) (t (t+1) \kappa+2)+1\rb +6 c_1  \hat{c} \lb t^4 (\kappa-2)-t^2 \kappa-1\rb e^{2 t} +3 \hat{c}^2 \lb (t-1) t ((t-1) t \kappa+2)+1\rb e^{4 t}}\,,
\ee
where
\be
t = |\gamma|\eta\,, \quad \kappa = 1-\frac{K}{\ga^2}\,, \quad \hat{c}=c_1+\pi c_2\,.
\ee
The effective equation of state is plotted in the right panel of figure \ref{eps2}. In our case, it turns out that for general $\nu$ and $c_1/c_2$, when $K>0$, the background fluid responsible for the bounce indeed typically respects the null energy condition. Thus we could reproduce this behaviour with e.g. a canonical scalar field. Obviously, with negative curvature one has to violate the null energy condition since $\rho_B \sim \h^2+K$. Except when $K>0$, the collapsing phase must be driven by a negative energy source. In the simplest form this is a negative cosmological constant, which however has to go through a phase transition to recover an observationally allowed vacuum density at later epochs; such scenarios have also been elaborated recently \cite{Biswas:2010si,Biswas:2009fv}. Here it is natural to consider the bounce to occur in a closed universe where the collapsing phase is driven by supercritical density of matter. To generate a bounce, this density needs to violate only the strong energy condition. In the following we find that the inflationary infrared divergences are absent in such models.
\newline
\newline
To close this section, we note that when $\nu$ is not a half-integer, the solutions for negative $\ga^2$ are written then in terms of the modified Bessel functions, and the solutions for imaginary $\ga^2$ in terms of Bessel functions. This is the opposite to the above cases in subsections \ref{pos} and \ref{neg}. Furthermore, with a positive nonminimal coupling $\xi>1/6$, the solutions once again switch their roles.

\section{Scalar propagator for quantum fluctuations}
\label{back_sec}

 Here we first derive the scalar propagator and evaluate it at the coincidence limit in \ref{pr_sec}. 
 The cases where this is due to infinities lost in the analytic continuation are identified and discarded (see figure \ref{IRplot}).
 We then renormalize the ultraviolet divergencies in a standard way by adding suitable counterterms in \ref{re_sec}. In section \ref{limit_sec} we show that the previous results are recovered in the appropriate limits.

\subsection{Infrared-regular sum over the modes}
\label{pr_sec}
\begin{figure}
\includegraphics[width=5.9cm]{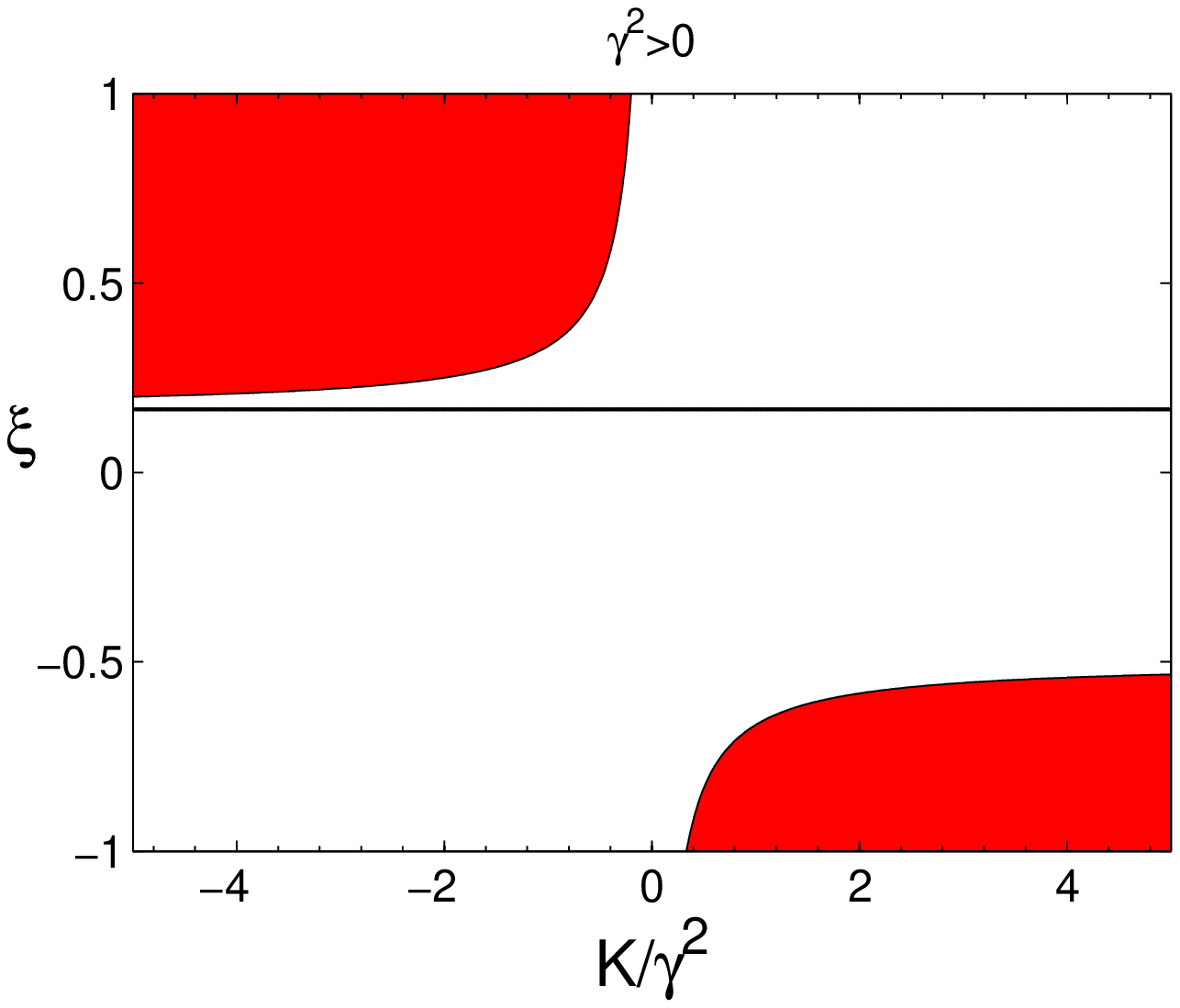}
\includegraphics[width=5.9cm]{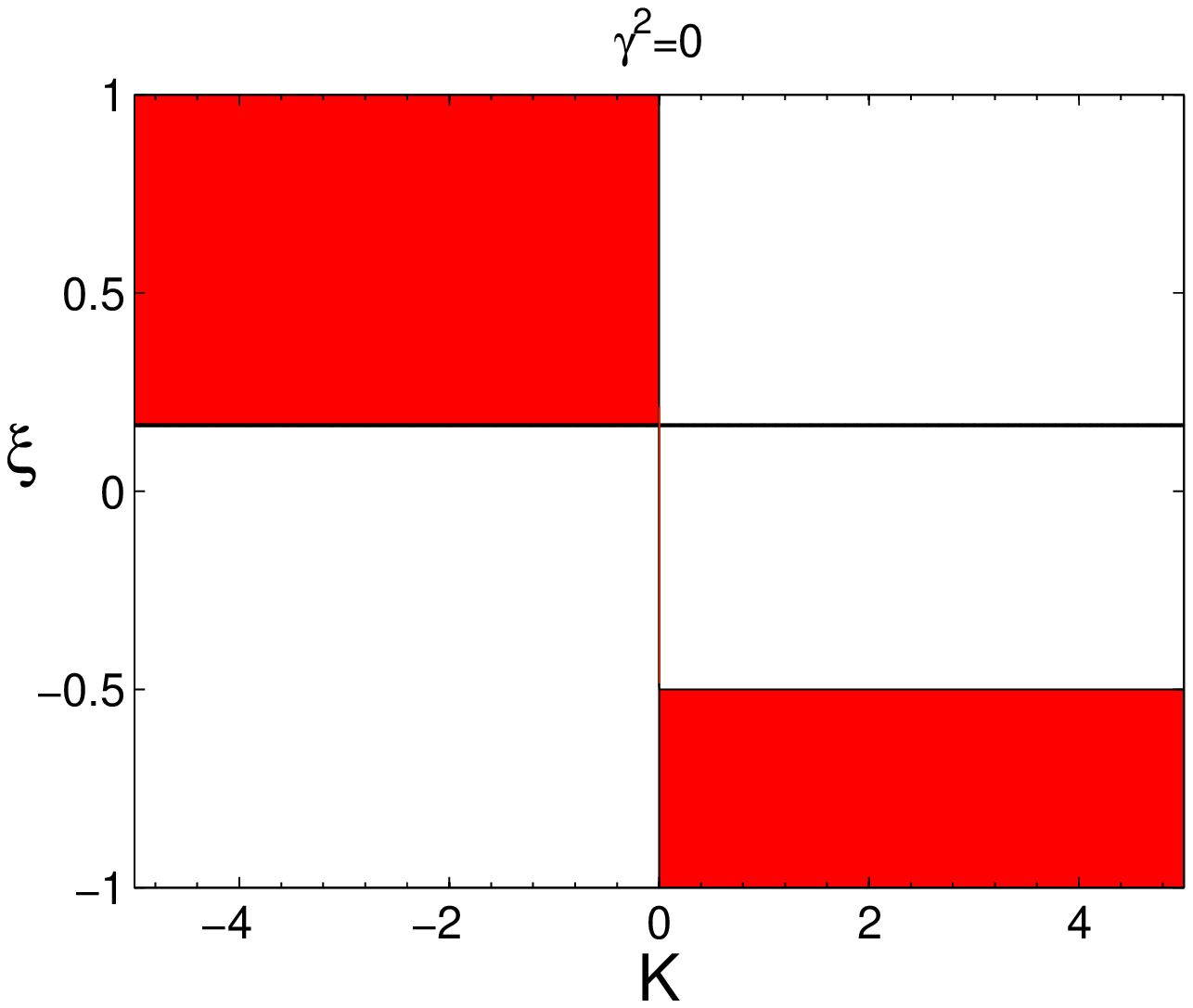}
\includegraphics[width=5.9cm]{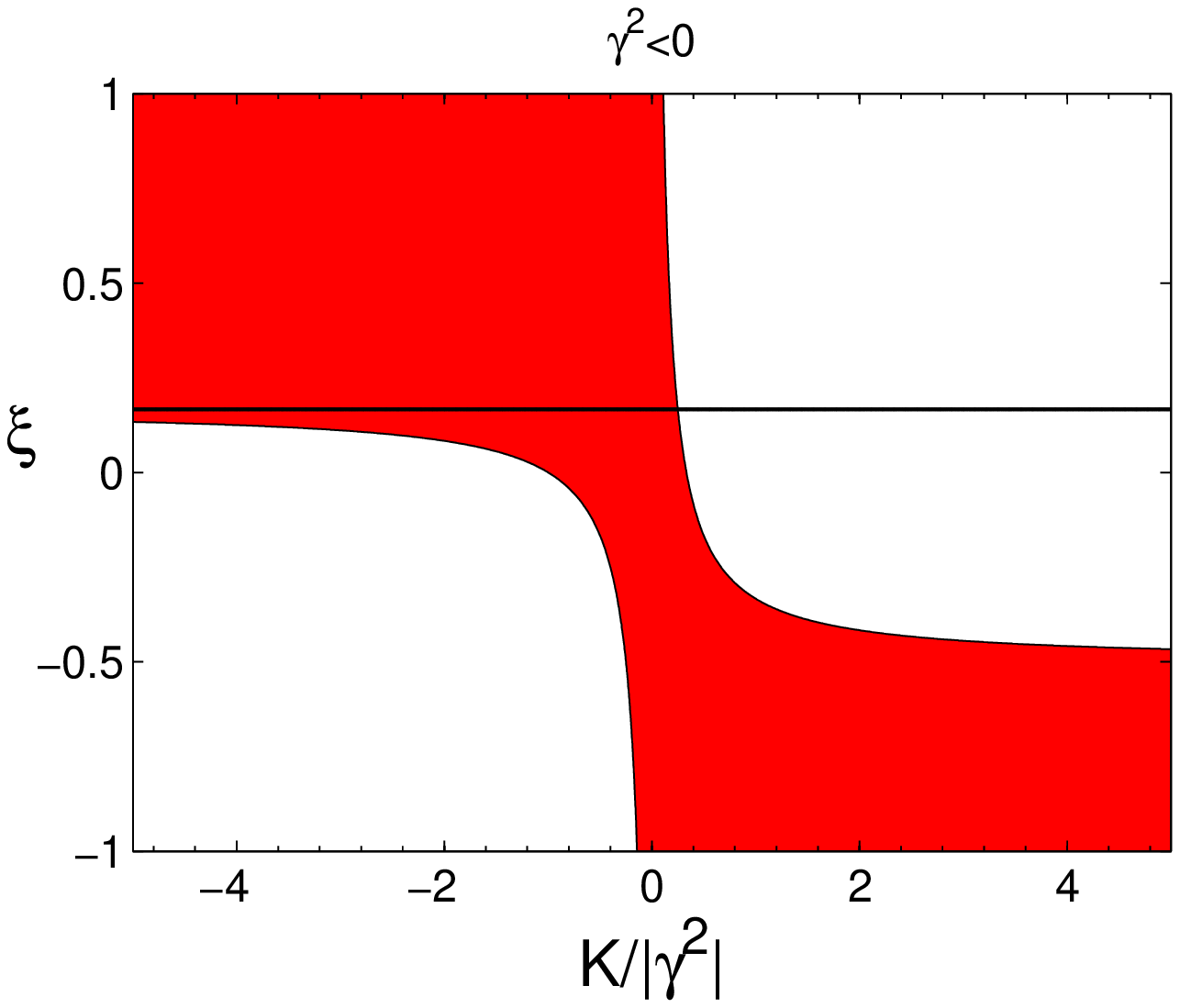}
\caption{\label{IRplot} The colored areas in the curvature-coupling plane are excluded because of the infrared divergence. The horizontal lines indicate the conformal coupling $\xi=\frac{1}{6}$.
{\bf Left}: The cases $\gamma^2>0$ (curvature in units of $\gamma^2$) whose background evolutions were discussed in \ref{pos}. The minimally coupled models $\xi=0$ are always regular. {\bf Middle}: The case $\gamma=0$, which corresponds to constant deceleration spacetimes. Note that the flat models $K=0$ are excluded.
{\bf Right}: The cases $\gamma^2<0$ (curvature in units of $-\gamma^2$) whose background evolutions were discussed in \ref{pos} and \ref{bounce}. The minimally coupled models $\xi=0$ require $K<\gamma^2$ or $K>\frac{1}{3}|\gamma|^2$.}
\end{figure}

The solution to the mode equation (\ref{p_evol}) with the ansatz (\ref{ansatz}) can be written in terms of the Hankel functions,
\be \label{psi}
\psi_k(\eta) = a^{1-\frac{D}{2}}(\eta)\sqrt{\frac{\pi|\eta|}{4}}\left[\al H_\nu^{(1)}(\sqrt{k^2 + \ga^2_K}|\eta|) + \bt H_\nu^{(2)}(\sqrt{k^2 +
\ga^2_K}|\eta|)\right]\,,
\ee
where we have defined
\be \label{gak}
\ga^2_K = \ga^2 - \left[\frac{1}{4}(D-2) - (D-1)\xi\right](D-2) K\,,
\ee
 which plays the role of a shifted $\ga^2$ in the presence of curvature. While $\ga^2$ specifies the evolution of the background, it is the $\ga_K^2$ which determines the properties of the fluctuations.
The normalization from the Wronskian (\ref{wronskian}) is such that $|\al|^2-|\bt|^2=1$.
The Bunch-Davies vacuum corresponds to the choice $\al=1$, $\bt=0$, which we make here. The time-ordered Feynman propagator, obeying Eq. (\ref{eom}), for the vacuum state $|\Omega\rangle$ is
\ba \label{propagator}
i\Delta(x;x')& = & \langle\Omega|\theta(\eta-\eta')\Phi(x)\Phi(x') + \theta(\eta'-\eta)\Phi(x')\Phi(x)|\Omega \rangle \nonumber \\
             & = & \int\frac{d^{D-1}k}{(2\pi)^{D-1}}Q^*({\bf k},{\bf x})Q({\bf k},{\bf x}')
             \left[ \theta(\eta-\eta')\psi(\eta)\psi^*(\eta') + \theta(\eta'-\eta)\psi(\eta')\psi^*(\eta)\right] \,.
\ea
In the second line we recalled the expansion (\ref{phi}) and used the definition of the vacuum $b_\bk|\Omega\rangle=0$. In the case of closed universe, one should, more rigorously, replace the integral by a sum over the modes as mentioned in appendix \ref{k_app}.
\newline
\newline
We shall need the propagator at the coincident limit. 
At coincidence the mode functions $Q({\bf k},{\bf x})$ reduce to plane waves.
After performing the angular integral in (\ref{propagator}) and inserting the solution (\ref{psi}) we get
\be \label{prop2}
i\Delta(x;x)  =  \frac{a^{2-D}|\eta|}{2^D \pi^{\frac{D-3}{2}}\Ga(\frac{D-1}{2})}\int_C^\infty dk k^{D-2} H_\nu^{(1)}(\sqrt{k^2 + \ga^2_K}|\eta|) H_\nu^{(2)}(\sqrt{k^2 + \ga^2_K}|\eta|)\,,
\ee
where $C$ is the cut-off that equals $D\sqrt{K}/2$ when $K>0$ and vanishes $C=0$ when $K<0$.
At this point we assume that $\nu$ is a half-integer. Then we may employ the finite series presentation \cite{gr}\footnote{The $1/2$-terms in the Gamma-factorials
have a different sign in Ref. \cite{gr} 4th edition in Eq. (8.466), but Eq. (8.451) is correct.}
\be \label{hankel}
H^{(1)}_\nu(z) = \sqrt{\frac{2}{\pi z}}i^{-(\nu+\frac{1}{2})}e^{iz}\sum_{n=0}^{\nu-\frac{1}{2}}\frac{\Ga(\nu+\frac{1}{2}+n)}{n!\Ga(\nu+\frac{1}{2}-n)} (-2iz)^{-n}\,,
\ee
and obtain
\be
i\Delta(x;x) = \frac{a^{2-D}}{2^{D-1} \pi^{\frac{D-1}{2}}\Ga(\frac{D-1}{2})}\int_C^\infty dk k^{D-2}\sum_{n,m=0}^{\nu-\frac{1}{2}}
\frac{i^{m-n}\Ga(\nu+\frac{1}{2}+n)\Ga(\nu+\frac{1}{2}+m)}{n!m!\Ga(\nu+\frac{1}{2}-n)\Ga(\nu+\frac{1}{2}-m)}
\frac{(k^2+\gamma_K^2)^{-\frac{1}{2}(1+n+m)}}{(2|\eta|)^{n+m}}\,.
\ee
The benefit is that the integral can be performed analytically. When the radius of convergence is nonzero, it can be shown that the result applies well for general $\nu$ \cite{Janssen:2009nz}, however in the present case the analytic extension seems more difficult. From the form of the integral we see that divergence is expected at the upper limit when $D-2 > n+m$, and that terms in the sum with odd $n+m$ vanish. In fact the double sum can be reorganized into a single sum as
\ba
i\Delta(x;x) & = & \frac{a^{2-D}}{2^{D-1}\pi^{\frac{D}{2}}\Ga(\frac{D-1}{2})}\int_C^\infty dk k^{D-2}\sum_{n=0}^{\nu-\frac{1}{2}}
\frac{\Ga(\frac{1}{2}+n)\Ga(2\nu+1+2n)\Ga(\nu+1-n)}{n!\Ga(2\nu+1-2n)\Ga(\nu+1+n)}
\frac{(k^2+\gamma_K^2)^{-\frac{1}{2}-n}}{(4\eta)^{2n}} \nonumber \\
& = & \frac{a^{2-D}}{2^{D-1}\pi^{\frac{D}{2}}\Ga(\frac{D-1}{2})}\int_C^\infty dk k^{D-2}\sum_{n=0}^{\nu-\frac{1}{2}}
\frac{\Ga(\frac{1}{2}+n)\Ga(\nu+\frac{1}{2}+n)}{n!\Ga(\nu+\frac{1}{2}-n)}
\frac{(k^2+\gamma_K^2)^{-\frac{1}{2}-n}}{\eta^{2n}}\,. \label{single}
\ea
In the second line we used the Legendre duplication formula.
We note that $\ga_K^2>-C^2$ is needed for the regularity at the lower limit of the integral. However, the results turn out to be analytically extendable to arbitrary imaginary $\ga_K$, but this is due to the automatic subtraction at play in the dimensional regularization\footnote{The incorrect use of dimensional regularization has been noted to hide power-law divergences in constant-$\epsilon$ spacetimes \cite{Janssen:2008px} and the breaking of de Sitter invariance of propagators in $\epsilon=0$ spacetimes \cite{Miao:2010vs}.}. This is why we discard the $\ga_K^2<-C^2$ cases and can restrict to real $\ga_K$ in flat and open universes and consider negative $\ga_K^2$ possible only in closed universes. The allowed spacetimes we are left with are then the following for the minimally coupled case:
\begin{itemize}
\item $\GG>0$: This case is always regular.
\item $\GG=0$: Spacetimes with constant deceleration are regular when the spatial sections are curved, $K\neq 0$.
\item $\GG<0$: This case is regular for open universes with $|K|>\frac{4|\GG|}{(D-2)^2}$ or closed universes with $K>\frac{|\GG|}{D-1}$.
\end{itemize}
These are the cases in which the integral in (\ref{single}) yields finite results without hiding the infrared power-law divergences by dimensional regularization.
The regularity conditions in the presence of a nonminimal coupling are illustrated in figure \ref{IRplot}.
Here we see that it is the combination of the background evolution, given by $\ga$, and the curvature, given by $K$, which determines the physically reasonable cases. We also note that both positive or negative spatial curvature can regulate the infrared, but by a different mechanisms. Negative $K$ contributes to the evolution of the fluctuation modes, mimicking a real $\ga$, and can thus render these modes less singular at very large scales. On the other hand, with a positive $K$ the very large scale modes are simply absent, because they cannot be excited in a closed universe.
\newline
\newline
Keeping in mind the restriction to the above cases, we perform the integral in (\ref{single}) to obtain
\ba \label{series1}
i\Delta(x;x) & = & \frac{C^{D-2}}{a^{D-2}2^{D-1} \pi^{\frac{D}{2}}\Ga(\frac{D-1}{2})}\sum_{n=0}^{\nu-\frac{1}{2}}
\frac{\Ga(\frac{1}{2}+n)\Ga(\nu+\frac{1}{2}+n)}{n!(2(1+n)-D)\Ga(\nu+\frac{1}{2}-n)}
\nonumber \\ & \times & {}_2F_1\lp 1-\frac{D}{2}+n,\frac{1}{2}+n;2-\frac{D}{2}+n; -\frac{\ga_K^2}{C^2}\rp \frac{1}{(C\eta)^{2n}}\,,
\ea
in terms of the hypergeometric functions with arguments proportional to square of the inverse cut-off.
The result diverges in the limit  $\ga_K \rightarrow 0$ if $\nu>\frac{1}{2}$. Thus by introducing $\gamma_K$ we have indeed regularized the infrared. However, an ultraviolet divergence remains. We will deal with that in the following subsection.
\newline
\newline
First we note that it is useful to rewrite this result as
\ba \label{series}
i\Delta(x;x) & = & \frac{\ga_K^{D-2}}{a^{D-2}2^D \pi^{\frac{D}{2}}}\sum_{n=0}^{\nu-\frac{1}{2}}
\frac{\Ga(1-\frac{D}{2}+n)\Ga(\nu+\frac{1}{2}+n)}{n!\Ga(\nu+\frac{1}{2}-n)}\frac{1}{(\ga_K\eta)^{2n}}
\nonumber \\ & \cdot & \left[1-\left( \frac{C}{\ga_K} \right)^{D-1}\frac{\Ga(\frac{1}{2}+n)}{\Ga(1-\frac{D}{2}+n)\Ga(\frac{D+1}{2})}
\times{}_2F_1\lp \frac{1}{2}(D-1),\frac{1}{2}+n;\frac{1}{2}(D+1);-\frac{C^2}{\ga_K^2}\rp\right]\,.
\ea
The first line is the limiting value of the above (\ref{series1}) as $C \rightarrow 0$, and is thus the complete result for flat and open universes. The second line deviates from unity in the presence of a nonzero cut-off $C$, and in particular it takes into account the absence of modes with $k<\frac{D}{2}\sqrt{K}$ in closed universes. Especially in the case of vanishing cut-off this expression is more practical than (\ref{series1}). However, we should stress that the Gauss hypergeometric functions appearing in (\ref{series}) have a branch cut at imaginary $\ga_K$ and in that case the use of the expression (\ref{series}) requires some extra care. The expression (\ref{series1}) is straightforwardly applicable in the regime $\ga^2_K/C>-1$ covering precisely the physically regulated cases. The expressions are equivalent, but for practical purposes (\ref{series1}) should be used when $C^2 \gg \ga_K^2$, and (\ref{series}) when $C^2 \ll \ga_K^2$ (and in particular when $C=0$).

\subsection{Regularizing the ultraviolet by counter-terms}
\label{re_sec}

Let us first look at the case (\ref{series}). 
The first two nonvanishing terms in the series (\ref{series}) contain the divergent piece:
\ba
i\Delta^{UV}(x;x) & = &
\frac{a^{2-D}}{2^{D+3}\pi^{\frac{D}{2}}}\frac{(\gamma_K/\mu)^{D-4}}{\eta^2}\Bigg\{\Ga(1-\frac{D}{2})\lb 8\gamma_K^2\eta^2+(D-2)\left(1-4\nu^2\right)\rb
\nonumber \\
& - & \lp\frac{C}{\ga_K}\rp^{D-1} \frac{\sqrt{\pi}}{\Gamma(\frac{1}{2}(D+1))} \Big[ 8 \ga_K^2 \eta^2 \times {}_2F_1\lp\frac{1}{2},\frac{1}{2}(D-1); \frac{1}{2}(D+1); -\frac{C^2}{\ga_K^2}\rp \nonumber \\ & + & (4\nu^2-1)\times {}_2F_1\lp\frac{3}{2},\frac{1}{2}(D-1); \frac{1}{2}(D+1); -\frac{C^2}{\ga_K^2}\rp\Big]\Bigg\}\,, \label{div2}
\ea
where $\mu$ is an arbitrary renormalization scale with the dimension of mass. The large-scale cut-off is not relevant for the ultraviolet divergence, and is thus contained $C=0$ part of the above propagator. This is the first line of (\ref{div2}). The hypergeometric functions in the two following lines are regular in $D=4$ and can be straightforwardly evaluated there.
This becomes, near $D=4$,
\ba \label{conver}
\mu^{4-D}i\Delta^{UV}(x;x) & = & \frac{1-4\nu^2+4(\ga_K\eta)^2}{32\pi^2a^2\eta^2(D-4)} + \frac{1-4\nu^2}{64\pi^2a^2\eta^2}\lb \gamma_E +
\log{\lp\frac{\ga_K^2}{4\pi a^2\mu^2}\rp} \rb
\nonumber \\
& - & \frac{\ga_K^2}{16\pi^2a^2}\lb 1-\gamma_E-\log{\lp\frac{\ga_K^2}{4\pi a^2\mu^2}\rp}+2\frac{K}{\ga_K^2}\lp1-5\xi\rp\rb
\nonumber \\
& - & \frac{1}{32a^2\pi^2 \eta^2}\lb \frac{C\lp 1-4\nu^2 + 4(\ga_K^2+C^2)\eta^2\rp}{\sqrt{\ga_K^2+C^2}} + \lp 1-4\nu^2+4\ga_K^2\eta^2\rp\arcsin\lp\frac{C}{\ga_K}\rp\rb
+ \mathcal{O}(D-4)\,.
\ea
We have multiplied the propagator by $\mu^{4-D}$ to obtain the correct mass dimension off $D=4$.
Here $\ga_K$ is evaluated in $D=4$. We denote the Euler-Mascheroni constant by $\ga_E$ and $\psi(x)=\frac{d}{dz}\log\lp\Gamma(z)\rp$ is the digamma function. The last line in (\ref{conver}) contains the contribution from the hypergeometric functions that vanishes when $C=0$.
In the minimal subtraction scheme, all but the first term in Eq. (\ref{conver}) contribute to the renormalized propagator.
One may write an equivalent expression either by directly expanding the first two terms in the series in (\ref{series1}) about $D=4$ or by using the previous
result\footnote{We note that in (\ref{conver}) an imaginary $\ga_K$ inside the logarithm would require an imaginary $\mu$ for consistency. This would spoil the unitarity of the counter term lagrangian (\ref{ct_lag}), which is one way to see that the $\ga_K^2<0$ cases are unphysical in flat and open universes. However, we can allow an imaginary $\ga_K$ in a closed universe. As argued above, the propagator (\ref{series1}) is then devoid of infrared singularities given $C>|\ga_K|$, and when expanded about $D=4$ in Eq. (\ref{conver1}), is also perfectly compatible with a real renormalization scale $\mu$. One should just note that in the case of imaginary $\gamma_K$, the different branch-cut should be chosen in Eqs. (\ref{series},\ref{conver}).}
(\ref{conver}). We obtain
\ba \label{conver1}
\frac{i\Delta^{UV}(x;x)}{\mu^{D-4}} & = & \frac{1 -4\nu^2+4(\ga_K\eta)^2}{32\pi^2\eta^2(D-4)} -  \frac{1-4\nu^2}{64\pi^2a^2\eta^2}\lb 2 - \ga_E -
\log{\lp\frac{C^2}{\pi\mu^2a^2}\rp}
- \frac{3\ga_K^2}{2C^2} + \frac{15\ga_K^4}{16C^4}\times{}_2{F}_2\lp \frac{7}{2},2;3,3;-\frac{\ga_K^2}{C^2}\rp
\rb \nonumber \\
 & - &  \frac{1}{8\pi^2a^2}\lb \lp 2-\ga_E-\log{\lp\frac{C^2}{\pi\mu^2a^2}\rp} \rp\frac{\ga_K^2}{2} + (1-5\xi)K
 +  C^2  - \frac{3\ga_K^4}{8C^2}\times{}_3{F}_2\lp \frac{5}{2},1,1;2,3;-\frac{\ga_K^2}{C^2}\rp \rb\,,
\ea
up to terms vanishing in $D=4$. This form is more convenient to use when $C$ is large compared to $|\ga_K|$.
\newline
\newline
We will be interested in computing the one-loop expectation value for the trace of the scalar field energy momentum tensor (\ref{emt}). The value of the trace (\ref{trace}) is given by, after using the Eq. (\ref{eom}) and the equation of motion $(\Box-\xi R)\Phi=0$,
\be \label{trace2}
\langle \Omega| T | \Omega \rangle = -\left(\frac{D-2}{4}-(D-1)\xi\right)\Box i \Delta(x;x)\,.
\ee
At this point we note that the ultraviolet divergence (\ref{conver}) is proportional to the scalar curvature (\ref{curvature}). This is what we expect, since the one-loop correction is known to be renormalizable with simple geometric counterterms.
Generally, we may consider the two quadratic geometric counterterms (as the Riemann squared in $D=4$ is a linear combination of these)
\be \label{ct_lag}
\mathcal{L}_{ct} = -\alpha_1 R^2 - \alpha_2 R_{\mu\nu}R^{\mu\nu}\,.
\ee
This lagrangian contributes to the trace as
\be \label{ctrace}
T_{ct} = \left[-4(D-1)\alpha_1 + \alpha_2D\right]\Box R + (D-4)\left(\alpha_1 R^2 + \alpha_2 R_{\mu\nu}R^{\mu\nu}\right)\,.
\ee
Thus the (infinite parts of the) coefficients that renormalize the theory are
\be
(D-4)(\al_1-\frac{1}{3}\al_2) = \frac{(1-6\xi)^2}{1152\pi^2} \mu^{D-4}\,.
\ee
As $\ga_K$ is related to infrared physics and doesn't enter into this expression, it coincides with the result in \cite{Janssen:2009nz}.
The total trace from the divergent part of the scalar field stress energy (\ref{trace2}) and the counter terms (\ref{ctrace}) becomes
\ba \label{diver}
\langle \Omega|T_{div} + T_{ct}| \Omega \rangle  & = &  \frac{\mu^{D-4}}{32\pi^2a^4}\left(\gamma_K^2+\frac{\frac{1}{4}-\nu^2}{\eta^2}\right)^2
 +  \frac{12(D-4)\al_2}{a^4}\left[2\h^4+3\h^2\h'+2(\h')^2-K(4\h^2+3\h'-4K)\right] \nonumber \\
 & + & \frac{24}{a^4}\lp 3\al_1^{fin}-\al_2^{fin}\rp\lp \h''' -6\h^2\h'-2(\h')^2+2\h'K\rp\,.
\ea
The second line contains the contributions from the possible finite parts of the coefficients $\al_1$ and $\al_2$. A simple choice for renormalization is to set
$\al_2$ and the finite parts of both coefficients to zero. The latter corresponds to the minimal subtraction scheme in dimensional regularization. We will use
this choice in the numerical examples that follow in section \ref{ba_sec}.

\subsection{de Sitter and quasi-de Sitter limits}
\label{limit_sec}

\subsubsection{de Sitter limit}

It is known for a long time that in a locally de Sitter space the propagator of a minimally coupled massless scalar at the coincidence limit contains a logarithmically growing part \cite{Vilenkin:1982wt,Starobinsky:1982ee,Linde:1982uu}. To check the consistency of our results with the previous literature, we consider the asymptotic limit when $\nu=3/2$ and
\be \label{dslimit}
\h/a \rightarrow \frac{c_2}{4}\sqrt{\frac{\pi}{2}}|\ga| \equiv H_0 \,, \quad \eta \rightarrow -\frac{1}{aH_0}\,.
\ee
At small enough $|\eta|$, i.e. after a sufficient inflationary expansion, this becomes an excellent approximation.
Setting the cut-off to zero, the propagator (\ref{series}) becomes, near $D=4$,
\ba \label{dsprop}
\mu^{4-D}i\Delta(x;x) & = & \left(\frac{H_{0}}{4\pi}\right)^2\left[-\frac{4+2(\ga_K\eta)^2}{D-4}+2\log{a^2} - 2\ga_E  - 2\log{\lp \frac{\zeta}{4\pi^2\mu^2}\rp}
\right] \nonumber \\
& - & \frac{1}{16\pi^3a^2}\lb  2\lp \log{\lp \frac{\zeta}{4\pi a^2\mu^2}\rp} - 1 + \ga_E\rp\zeta + \pi K\rb\,,
\ea
where the constant $\zeta$ is defined as
\be
\zeta = \frac{c_2^2H_0^2}{16} - \pi K\,.
\ee
 Eq. (\ref{dsprop}) agrees with e.g. the result derived in Ref. \cite{Onemli:2002hr}. In the first line, the prefactor is the familiar square of the Hubble rate per $4\pi$, then we have a divergence, and the secular growth which is logarithmic in the scale factor. The following constant depends on the renormalization scale and other parameters. In the case of nonzero cut-off, this constant would have a different form. The second line represents a correction which decays as $1/a^2$, and can be neglected at sufficiently late times (the constant coefficients would again be modified if there was a cut-off). All such redshifting terms can be absorbed into a suitable modification of the initial state \cite{Kahya:2009sz}. Our results consistently generalize the previous coincident de Sitter propagators into the case where the initial (Bunch-Davies) state does not have to be set in the exact de Sitter background.

\subsubsection{Quasi de Sitter limit}
\label{qds}

We have essentially three parameters which determine the scale factor expansion: $\ga$ and $\nu$ specify the evolution equation (\ref{a_evol}) and the ratio
$c_2/c_1$ of the constants in the solutions discussed in section \ref{background} specify a particular solution to this equation. Qualitatively different classes
of spacetimes emerge when the sign of $\ga^2$ is changed or one of the coefficients $c_1$ or $c_2$ is switched on or off, as detailed in section \ref{background}.
However, we were able to carry out the computations analytically only for a discrete set of $\nu$-values. This seems to restrict us to consider
only a limited set of inflationary models. In particular, to study spacetimes which become asymptotically infinitesimally close to de Sitter would require to treat $\nu$ as a continuous parameter. Analytical extension of the result (\ref{series1}) to arbitrary $\nu$ though does not seem straightforward as the coefficients in the series diverge factorially.
\newline
\newline
Instead, we can consider models with arbitrary slow-roll parameters in the presence of a non-minimal coupling. By performing a sort of rotation in the parameter space, we can extend the results continuously to any value of effective $\nu$ and thus to any value of slow-roll parameter $\epsilon$.
Recall the scaling (\ref{xi_scale}), which relates the $\nu_\xi$ that determines the background expansion, to the parameter $\nu$ that governs the behavior of the
scalar field mode functions in (\ref{psi}). Using this, one readily deduces that for a given $\nu$, the slow-roll parameter will have the value $\epsilon$ at
asymptotically late times if the coupling is set to \be \label{epsilonxi}
\xi=\frac{(3-\epsilon)^2-4\nu^2(1-\epsilon)^2}{24(2-\epsilon)}\,.
\ee
This means also that we have an asymptotic de Sitter expansion for any $\nu$ when $\xi=(9-4\nu^2)/48$. For $\nu=\frac{5}{2}$ this already requires a rather large coupling $\xi=-1/3$. It seems thus most natural to consider quasi-de Sitter spaces expanding about $\nu=3/2$, since then we need only an infinitesimal coupling $\xi$ to have $\epsilon$ deviating infinitesimally from zero: $\xi = \frac{1}{4}\epsilon - \frac{1}{6}\sum_{n=2}^\infty(\frac{\epsilon}{2})^n$.
\newline
\newline
This seems to suggest that the result (\ref{dsprop}) is robust to small deviations from exactly exponential expansion. A nonzero $\epsilon$, achieved by just adjusting $\xi\approx \epsilon/4$, nothing but slightly changes the coefficients of the decaying terms in the propagator, whereas the leading logarithm behavior is unaffected.
If $\epsilon$ is constant, the expansion is described by
\be \label{eta}
-(1-\epsilon)\h=\frac{1}{\eta}\,.
\ee
Plugging this into (\ref{conver}), we obtain
\be
\mu^{4-D}i\Delta(x;x) \rightarrow \lp\frac{\h(1-\epsilon)}{4\pi a}\rp^2\lb -\frac{4}{D-4} + 2\log\lp \frac{4\pi\mu^2
a^2}{|\ga^2-(1-\frac{1}{6}\xi)K|}\rp\rb\,,
\ee
where $\xi$ is given by (\ref{epsilonxi}).
If $\epsilon$ is positive, the prefactor decays, and in super-inflating case, $\epsilon<0$, it grows. 


\section{Backreaction}
\label{ba_sec}

In this section we first compute the general expression for the one-loop contribution to the trace in \ref{1loop}. This is used to analyze the backreaction in inflationary
cosmologies. In \ref{inflation} we compute the asymptotic contribution to the effective energy density. In \ref{br_bounce} we consider models with contracting and
bouncing geometries. 

\subsection{One-loop contribution to the trace}
\label{1loop}

\begin{figure}
\includegraphics[width=8.75cm]{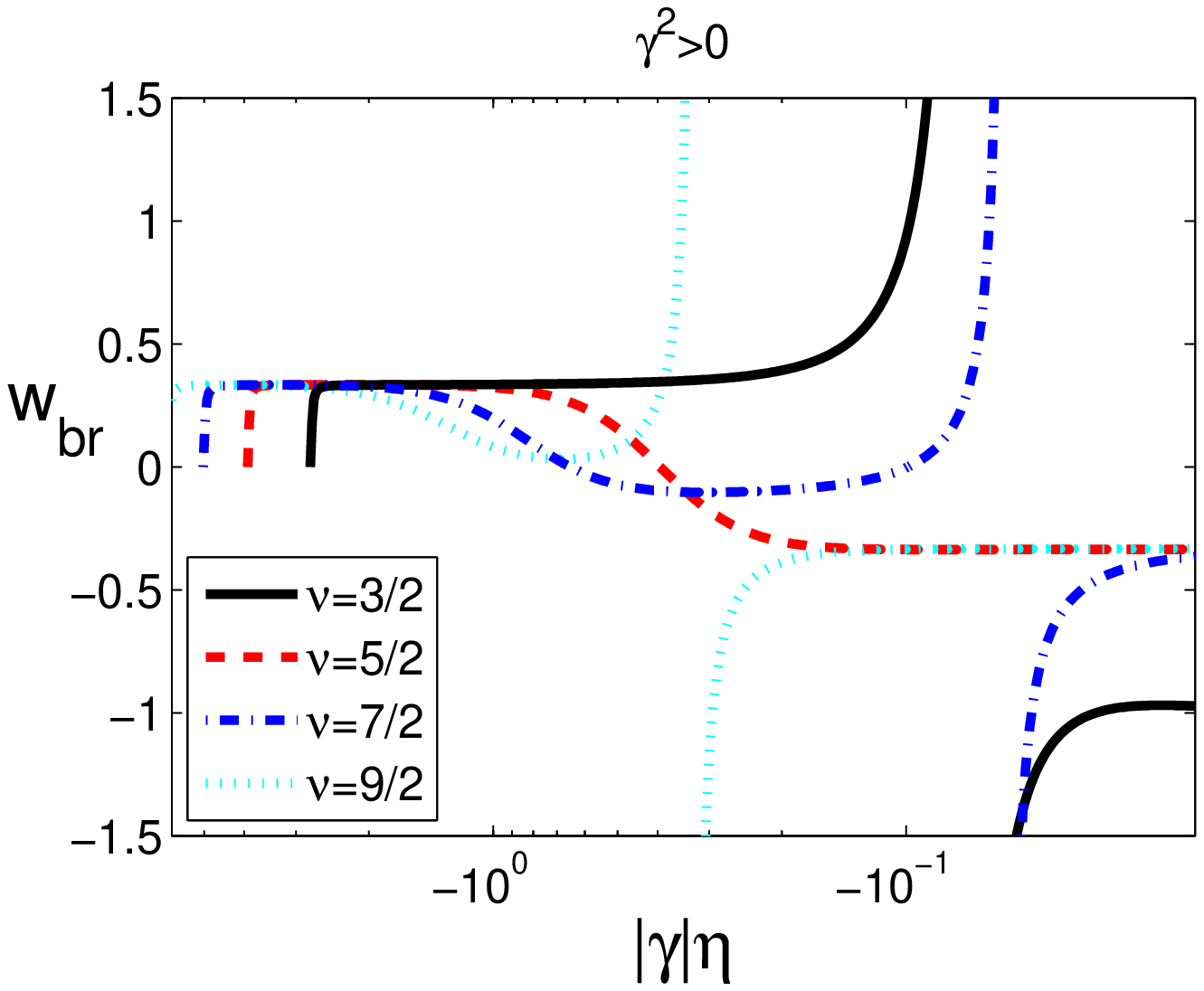}
\includegraphics[width=8.75cm]{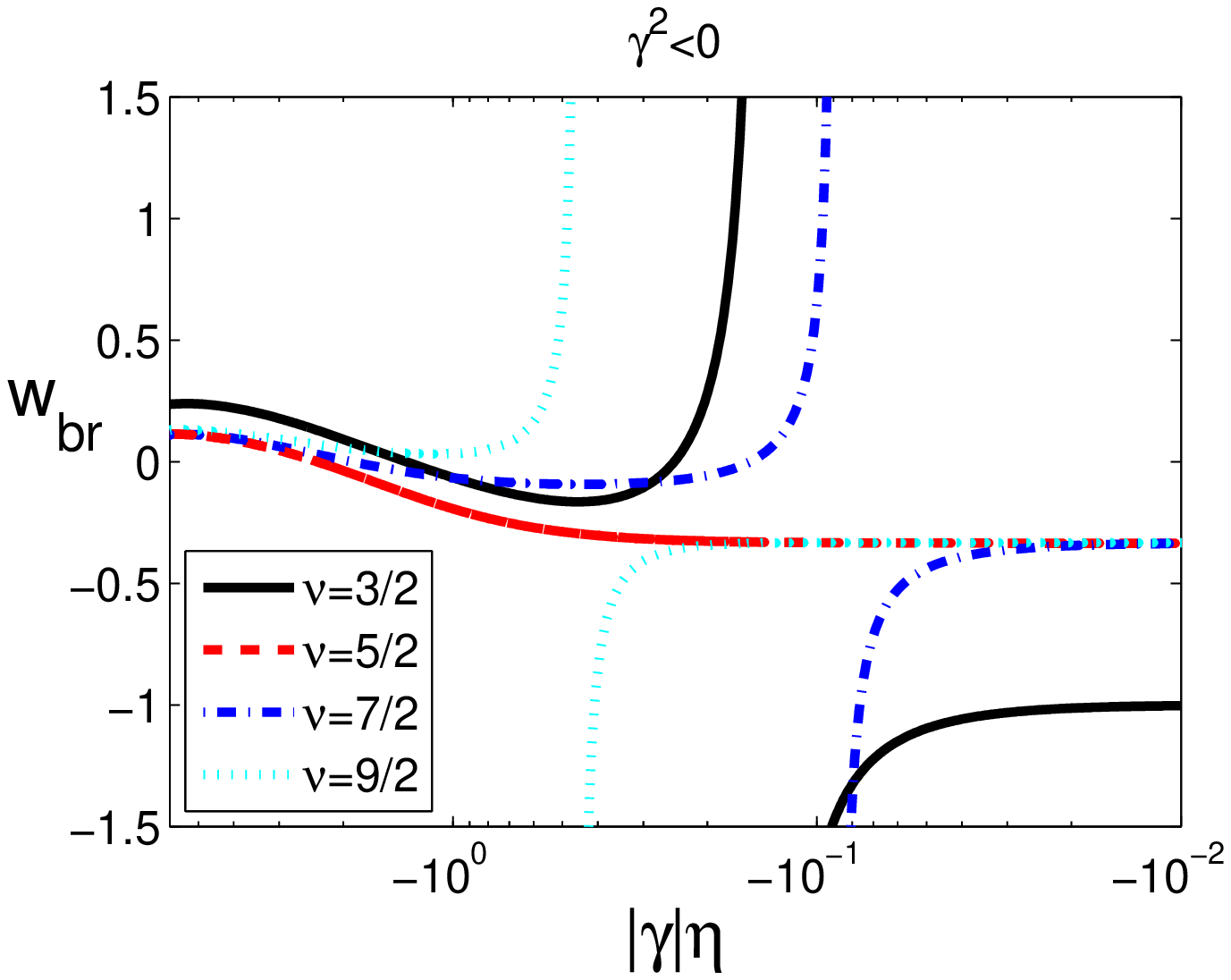}
\caption{\label{br0}
Evolution of the effective equation of state $w_{br}$ of the backreaction energy density as a function of conformal time in units of $|\ga|$. The backgrounds of
these models were displayed in figure \ref{eps1}.
Now $|\ga^2_K|=|\ga^2|=\mu$. 
{\bf Left}: The case of real $\ga$, where $K=0$. {\bf Right}: The case of imaginary $\ga$, corresponding to negative curvature $K=2\ga^2$.
In both cases, the backreaction density scales as radiation in the early time. The effective equation of state typically diverges, as the backreaction energy density changes its sign from positive to negative. Then the scaling approaches a constant again, which is $w_{br} \rightarrow -1$ when $\nu=3/2$ and $w_{br} \rightarrow -1/3$ for the other examples.}
\end{figure}

\begin{figure}
\includegraphics[width=8.75cm]{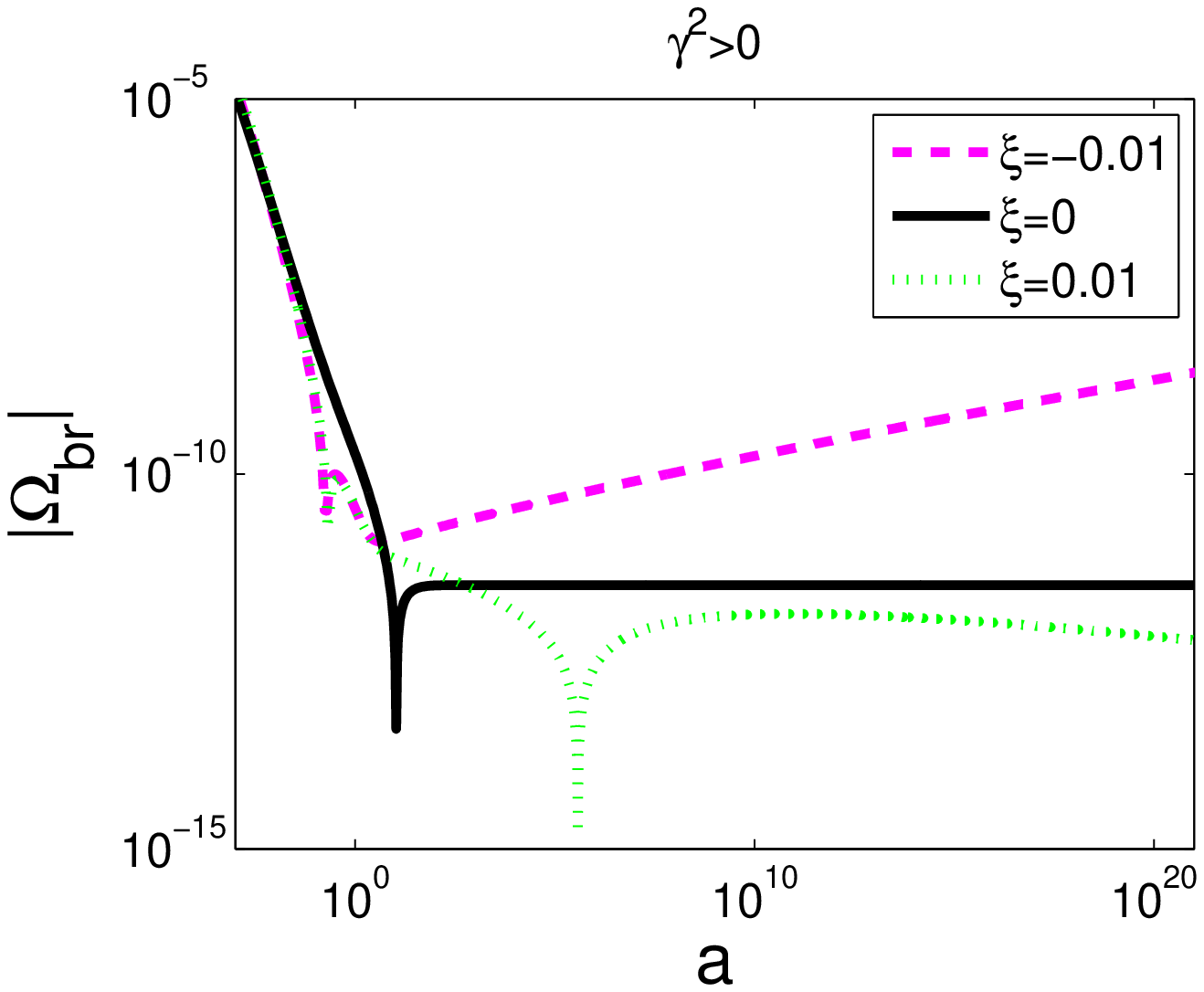}
\includegraphics[width=8.75cm]{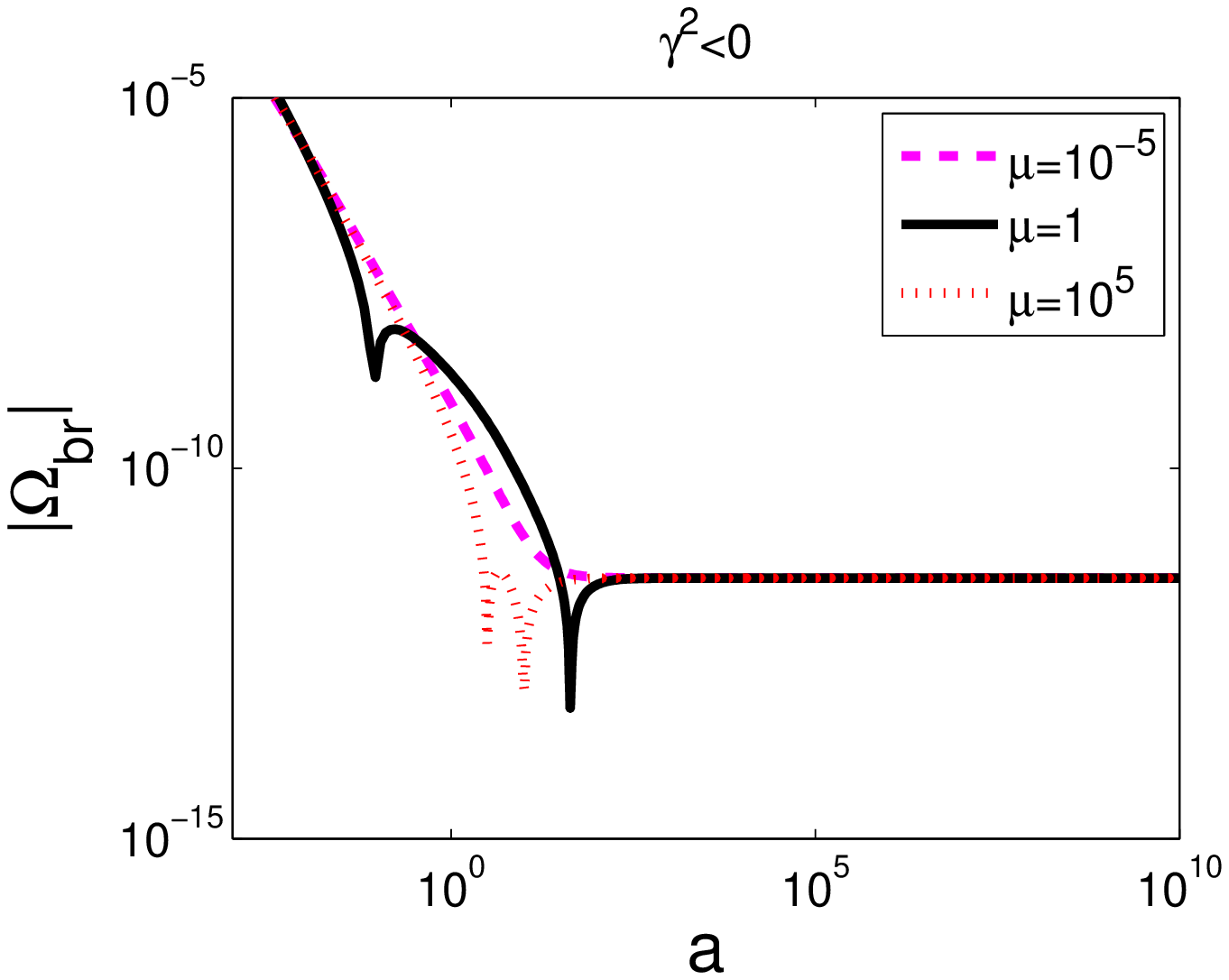}
\caption{\label{br1}Evolution of the fractional contribution to the energy density, $|\Omega_{br}|$ as defined in (\ref{br}), from the backreaction as a function of the e-folding time in flat models with $\nu=3/2$.
{\bf Left}: In the case of nonminimal coupling, the scaling of the quantum energy density is modified. When the coupling is negative, the backreaction grows.
{\bf Right}: Examples with different choices of $\mu$ when $K=2\ga^2$. As the renormalization scale appears only in the subleading terms, it does not affect the asymptotic results.
}
\end{figure}

To compute the expectation value of the trace (\ref{trace}), we have to act with the box on the regular part of (\ref{conver}) and add the contribution from the regularization (\ref{diver}). Finally, we must collect the remaining terms from (\ref{series1}). The result is, setting $D=4$,
\ba \label{tback}
\langle \Omega |T|\Omega\rangle & = & -\frac{1}{32\pi^2a^4\eta^4}\Bigg[ (1-4\nu^2)\lp 3k_1 + 2(1+k_1)\h\eta \rp +\lp 2\h^2-(1+k_1)\h'\rp\eta^2 + 4\lp 2\h^2-(1+k_2)\h'\rp \ga_K^2\eta^4  \nonumber \\ & + & \lb (1-4\nu^2)\lp 3+2\h\eta-\h'\eta^2 \rp -4\h' \ga_K^2\eta^4 \rb \log{a^2}\Bigg] \nonumber \\
 & - &   \frac{\ga_K^2}{8\pi^2 a^4} \sum_{n=2}^{\nu-\frac{1}{2}}\frac{\hat{\Gamma}_n\Gamma(\nu+\frac{1}{2}+n)}{(n-1)\Gamma(\nu+\frac{1}{2}-n)(\ga_K\eta)^{2n}}\lp \frac{2n+1}{\eta^2}+\frac{2\h}{\eta}-\frac{\h'}{n}\rp \nonumber \\
& + &   \frac{\tilde{\al}^{inf}}{a^4}\left[2\h^4+3\h^2\h'+2(\h')^2-K(4\h^2+3\h'-4K)\right]  +  \frac{\tilde{\al}^{fin}}{a^4}\lp \h''' -6\h^2\h'-2(\h')^2+2\h'K\rp\,.
\ea
We have used shorthand notation for the dimensionless constants $k_1$ and $k_2$. In flat and open universes they are
\be \label{firstdef}
\left. \begin{array}{ccc}
k_1   & = & \ga_E+\log{\lp\frac{\ga_K^2}{4\pi\mu^2}\rp} \\
k_2   & = &  -1 +\ga_E+\log{\lp\frac{\ga_K^2}{4\pi\mu^2}\rp} - 2(1-5\xi)\frac{K}{\ga_K^2}
\end{array}\right\} \text{when} \,\, K \le 0\,.
\ee
In closed universes the appropriate constant are
\be
\left. \begin{array}{ccc}
k_1 & = & -2 + \ga_E + \log{\lp\frac{4K}{\pi\mu^2}\rp} + \frac{3\ga_K^2}{8K} - \frac{15\ga_K^4}{64K^2}\times{}_2{F}_2\lp \frac{7}{2},2;3,3;-\frac{\ga_K^2}{4K}\rp \\ 
k_2 & = &  -\frac{5}{2} + \ga_E+\log{\lp\frac{4K}{\pi\mu^2}\rp} - 2(3-5\xi)\frac{K}{\ga_K^2} + \frac{3\ga_K^2}{32 K}\times{}_3{F}_2\lp \frac{5}{2},1,1;2,3;-\frac{\ga_K^2}{4K}\rp
\end{array}\right\} \text{when} \,\, K > 0\,.
\ee
We also introduced the coefficients $\hat{\Gamma}_n$, defined as
\be
\hat{\Gamma}_n = \left\{ \begin{array}{ccc}   1 & \text{when} & K \le 0\,, \label{gamman1} \\
  \lp\frac{\ga_K^2}{4K}\rp^{n-1}\frac{2n\Gamma(\frac{1}{2}+n)}{\sqrt{\pi}}\times{}_2F_1\lp n-1,\frac{1}{2}+n; n; -\frac{\ga_K^2}{4K}\rp
&  \text{when} & K > 0\,. \end{array}\right.
\ee
Finally, in the last line of (\ref{tback}) the contribution from counterterms is given as
\ba
\tilde{\al}^{inf} & = & 12(D-4)\al_2\,, \\
\tilde{\al}^{fin} & = & 24(3\al_1^{fin}-\al_2^{fin})\,, \label{lastdef}
\ea
in terms of the coefficients specified in the lagrangian (\ref{ct_lag}). As expected, the result (\ref{tback}) diverges when $\ga_K\rightarrow 0$. The divergence
is logarithmic for $\nu=3/2$, and power-law $\sim \ga_K^{\frac{1}{2}-\nu}$ for $\nu>\frac{3}{2}$.
\newline
\newline
We can obtain the backreaction energy density by integrating (\ref{tback}), and then immediately deduce also the backreaction pressure. This is because we know that the effective quantum fluid must obey a continuity equation similar to (\ref{continuityB}), and furthermore the trace is $T=-\rho_{br}+3p_{br}$, the lower index $_{br}$ referring to the backreaction. Hence we have that
\be \label{continuity}
(a^4\rho_{br})'= -a^4\h T\,.
\ee
The integration constant corresponds to a freedom to add a radiation-like component. In the following we show numerical results for the integration.
In all the numerical examples presented here, we use the minimal subtraction scheme and set $\al_2=0$.
Unless otherwise specified, $\xi=0$, $|\ga|=|\ga_K|=\mu$,
$\nu=3/2$ and $K=0$. In realistic models, there is a large hierarchy between $K$ and the scale of inflation, $H_0$ in Eq. (\ref{dslimit}). In units of the Planck
scale, a GUT-scale inflation for example corresponds to $\sqrt{8\pi G_N} H_0 \sim 10^{-5}$, where $G_N$ is the Newton's constant, whereas the present
observational bounds on the curvature radius of the universe, quoted in the first paragraph of this paper, translate into $\sqrt{8\pi G_N |K|} \lesssim 10^{-43}$.
It is reasonable to assume that the parameter $\ga^2$ and thus also $\ga_K^2$ are roughly on the order of $K$. The scale factor can be arbitrarily rescaled, but we once fix $\h=H_0$ at $a=1$, this corresponds to the choice
\be
|c_2| = \frac{\sqrt{2}}{\Gamma(\nu)}\lp \frac{(2\nu-1)|\ga|}{4H_0} \rp^{\nu-\frac{1}{2}}\,,
\ee
in equation (\ref{bes}) or (\ref{modbes}) in section \ref{background}. In the example plots here the numerical ratio is $|\ga|/H_0 = 10^{-35}$.
\newline
\newline
We describe the relative significance and the behavior of the backreaction energy density by defining the dimensionless quantities
\be \label{br}
\Omega_{br} = \frac{8 \pi G_N a^2\rho_{br}}{3\h^2}\,, \quad w_{br} = \frac{p_{br}}{\rho_{br}}\,.
\ee
The results in terms of $\Omega_{br}$ are plotted for an asymptotically de Sitter model in Figure \ref{br1}. The effective equation of state is plotted in the figure \ref{br0}.
The background expansions in this figure are precisely the ones depicted in figure \ref{eps1}. We see that for both real and imaginary $\ga$, the evolutions have
some qualitative similarities. Namely, initially the density scales as radiation ($w_{br} \approx \frac{1}{3}$), and in the future will scale tend to a negative
constant, which in the case $\nu=\frac{3}{2}$ is $w_{br}=-1$. In the other cases, the backreaction scales asymptotically like curvature, $w_{br}=-1/3$. Note that the formally defined effective equation of state can
diverge when the backreaction energy density changes its sign. The asymptotic scaling we observe in the plots is not obviously seen from (\ref{tback}) even in
the minimally coupled case, as the would-be-leading order terms systematically cancel at late times. The asymptotically leading corrections will be discussed in
detail below in \ref{inflation}. Nevertheless, it holds qualitatively that the backreaction energy density decays if $0<\epsilon<1$, goes like a logarithm in de
Sitter case and grows if the universe is undergoing a phantom expansion with $\epsilon<0$.
\newline
\newline
The effects of varying the coupling $\xi$ or the renormalization scale $\mu$ in the case $\nu=\frac{3}{2}$ in flat and open models are shown in figure \ref{br1}. With a negative coupling $\xi$, the relative backreaction contribution grows as a power-law. We will see below that this is a generic feature, in agreement with Ref. \cite{Janssen:2009nz}. With a positive coupling, the quantum contribution decays. As already seen in figure \ref{br1}, in the nonminimally coupled case, $\Omega_{br}$ tends to a constant. This constant is determined below in (\ref{dS}). The renormalization scale affects only the transient dynamics. Typically the
contribution from $\Omega_{br}$ is positive in the early times, but becomes negative at the inflationary epoch. In particular the time(s) when the sign changes depends upon the renormalization scale.

\subsection{Asymptotic behavior at constant deceleration}
\label{inflation}

Assuming that the background has constant deceleration, it is straightforward to integrate (\ref{tback}) exactly to obtain the energy density. This assumption is always valid in the far future in the above described models\footnote{If taken to be valid at all times, this corresponds to setting $\ga=0$ in Eq. (\ref{a_evol}). As noted above in section \ref{pr_sec}, then one needs nonzero $K$ to regularize the infrared. This would correspond to a fine-tuned situation where the energy density of an effective background fluid is $8\pi G_N\rho_B = K/a^2 + 3H_0^2/a^{2\epsilon}$, the first term canceling exactly the effect of nonzero curvature to the expansion, and the second term driving the constant-deceleration expansion. In anisotropic cosmologies such cancelling results in isotropy of the cosmic microwave background radiation \cite{Carneiro:2001fz} and might be explained by a presence of a two-form field \cite{Koivisto:2010dr}.}. We use equation (\ref{eta}) and call $H_0=\h(a=1)$. To take properly into account a possible nonminimal coupling, we should replace $\nu \rightarrow \nu_\xi$ given by (\ref{xi_scale}) for the background equations, as discussed in \ref{qds}.
We obtain then
\ba \label{rho_br}
\rho_{br} & = & \frac{(2\nu_\xi-1)}{4096\pi^2\eta^4}\lp \frac{2H_0\eta}{1-2\nu_\xi}\rp^{4\nu_\xi-2}\Bigg\{(4\nu^2-1)\lb 23+72k_1-8(7+6k_1)\nu_\xi + 20\nu_\xi^2 \rb
\nonumber \\
& - & 8(2\nu_\xi-1)\lb 3(4\nu^2-1)(2\nu_\xi-3)-8(2\nu_\xi-1)(\ga_K\eta)^2\rb\log{\lp\frac{2H_0\eta}{1-2\nu_\xi}\rp}
-16 (2\nu_\xi-1)(7+4k_2-6\nu_\xi)(\ga_K\eta)^2\Bigg\}
 \nonumber \\
& - & \frac{\ga_K^2(2\nu_\xi-1)}{64\pi^2\eta^2}\lp \frac{2H_0 \eta}{1-2\nu_\xi}\rp^{4\nu_\xi-2}\sum_{n=2}^{\nu-\frac{1}{2}} \hat{\Gamma}_n \frac{(1+2n)(1-2\nu_\xi+2n)\Gamma(\nu+\frac{1}{2}+n)}{(n+1)n(n-1)\Gamma(\nu+\frac{1}{2}-n)(\ga_K\eta)^{2n}} \nonumber \\
& + &  \frac{(2\nu_\xi-1)^2}{32\pi^2\eta^4}\lp\frac{2 H_0\eta}{1-2\nu_\xi}\rp^{4\nu_\xi-2}\Bigg[
\tilde{\al}^{fin}\lp 11-4\nu_\xi(3\nu_\xi-2)+8K\eta^2\rp \nonumber \\
& + & \tilde{\al}^{inf}\lp 1-\nu_\xi-4\nu_\xi^3+4(1+4\nu_\xi)K\eta^2 + \frac{64}{2\nu_\xi-1}(K\eta^2)^2\log{\frac{\eta}{\eta_i}}\rp  \Bigg]\,.
\ea
The definitions of the constants $k_1$, $k_2$, $\hat{\Gamma}_n$, $\tilde{\al}^{inf}$ and $\tilde{\al}^{fin}$ are given in formulas (\ref{firstdef}-\ref{lastdef}).
\newline
\newline
Note that in the minimally coupled case $\nu_\xi=\nu$, the sum in the third line is truncated already at $n=\text{max}(2,\nu-\frac{3}{2})$ due to the coefficient which kills the $n=\nu-\frac{1}{2}$ term. This can be understood, since this term originates from the part of the propagator (\ref{series1}) that scales as $a^{-2}\eta^{1-2\nu}$. When Eq. (\ref{eta}) holds, this part is in fact a constant. Thus it doesn't contribute to the expectation value (\ref{trace2}) that is given by the derivatives of the propagator. One may expect that this would change in the case of a massive field, in which case then $\Omega_{br}$ would approach a constant. However, here we are restricted to the massless field. In the following we extract the leading order term in (\ref{rho_br}) at small $\eta$ in a few special cases.

\subsubsection{$\nu=\frac{3}{2}$}

Let us first set $\xi=0$. Then, in the case $\nu=3/2$ the background expansion is asymptotically described by $w_B=-1$. The
fractional backreaction energy density then becomes
\be \label{dS}
\Omega_{br} \rightarrow \frac{G_N H_0^2}{48\pi}\lb -5 + 16(\ga_K\eta)^2\log{a^2} + \mathcal{O}((\ga_K\eta)^2)
\rb\,.
\ee
Thus the quantum effect on the expansion tends to a constant which is completely independent of the model
parameters and the history leading to the de Sitter expansion. The result that effect is negative and magnitude
of the constant is proportional to the Hubble rate squared in units of the Planck mass was to be expected.
\newline
\newline
Let us then look at the quasi-de Sitter spaces and non-minimally coupled field. Then the above result generalizes
to
\ba \label{qdS}
\Omega_{br} & \rightarrow & -\frac{G_N\lb-(1-\epsilon)H_0 \eta\rb^{\frac{2}{1-\epsilon}}}{192\pi
(1-\epsilon)^3\eta^2}\Big[ 86\epsilon(2-\epsilon)\log{a^2} \nonumber \\ & + & (5-\epsilon)\lp
4-9\epsilon+12(1-\epsilon)\epsilon
k_1\rp +128\lp 7-\epsilon(7-2\epsilon)\rp\tilde{\al}^{inf} - 704(1-\epsilon)^3\tilde{\al}^{fin}  +
\mathcal{O}((\ga_K\eta)^2)\Big]\,.
\ea
Again the leading-order term in (\ref{qdS}) is parameter-independent.
We now infer that the relative backreaction density $|\Omega_{br}|$ grows, i.e. $w_{br}<w_B$, if $\xi < 0$. In
this case the background is super-inflating, $\epsilon<0$, but the backreaction energy density still grows faster
than $\rho_B$. This suggests that the quantum backreaction may hinder the Big Rip
that is implied for the super-accelerating background, in line with e.g. Ref. \cite{Nojiri:2005sx}.
For a positive coupling to curvature $\xi>0$, we have the opposite behavior: $\en>0$, meaning that
the background density decays, but $\rho_{br}$ will decay faster. Then $|\Omega_{br}|$ will decrease proportionally
to a negative power of the scale factor. This is due to the positive non-minimal coupling, without which the growth is only log-enhanced. The result (\ref{dS}) confirms that the well-known leading-logarithm
behavior in quasi-de Sitter spacetimes is robust. The density becomes negative, and thus the effect is to slow
down inflation, in accordance with, e.g. Ref. \cite{Tsamis:1996qq,Abramo:1997hu}.
One expects that the
$\epsilon\,$-suppression of the effect would not appear at higher loops, though in our case this remains to be
verified. The above deduced behaviors are illustrated in the figure \ref{br1}., where we monitor the evolution further
inside the inflationary epoch and plot the logarithm of $|\Omega_{br}|$. Qualitatively these results agree with those in Ref. \cite{Janssen:2009nz}, where the infrared was regulated by matching the fluctuation modes in a decelerating and accelerating flat FLRW with constant background equations of state.

\subsubsection{$\nu=\frac{5}{2}$}

 In the minimally coupled case, when $\nu = \frac{5}{2}$, $w_B \rightarrow -\frac{2}{3}$ which can be modelled by e.g. domain walls. In this case we find that
\ba
\Omega_{br} &  \rightarrow & \frac{G_N H_0^4\eta^2}{3072\pi}\lb -576\log{a^2} +9 - 54 k_1 -4096\tilde{\al}^{inf}+704 \tilde{\al}^{fin} + \mathcal{O}((\ga_K\eta)^2) \rb\,.
\ea
 It is obvious that the backreaction energy density dilutes more rapidly than the classical background energy density.  The term scaling as the $\rho_B$ drops out from (\ref{rho_br}) as explained above, and the dominant term will be the following one, enhanced by the logarithmic scaling. We have that $w_{br}=-\frac{1}{3}+\lp 3a\log{a}\rp^{-1}$, according with the figure \ref{br0}, where $w_{br}$ is seen to approach the scaling of curvature from above. 
\newline
\newline
In the case of nonminimal coupling, the  $n=2$ term in the sum appearing in (\ref{rho_br}) becomes nonzero and thus dominating at late times. The leading contribution is then
\ba
\Omega_{br}  & \rightarrow & -\frac{G_N\lb-(1-\epsilon)H_0 \eta\rb^{\frac{2}{1-\epsilon}}10(1-2\epsilon)}{\pi\ga_K^2\eta^4}\lb 1+ \mathcal{O}((\ga_K\eta)^2\log{a})\rb
\,\, \text{when}\, K \le 0\,, \nonumber \\
\Omega_{br}  & \rightarrow & -\frac{G_N\lb-(1-\epsilon)H_0 \eta\rb^{\frac{2}{1-\epsilon}}\eta^2 10\epsilon(2-\epsilon)}{3\pi\ga_K^2\eta^4}\lb  1- 8\lp 4+\frac{\ga_K^2}{K}\rp^{-\frac{3}{2}} + \mathcal{O}((\ga_K\eta)^2\log{a}) \rb
\,\, \text{when}\, K > 0\,.
\ea
We now deduce as above that if $\xi<0$  ($\xi>0$), the quantum backreaction grows (decays) like a power-law. Now the magnitude of the backreaction depends on the regularization parameter $\ga_K$. However, one deduces that it is still always negative for arbitrary curvature. There is no $\epsilon$-suppressed $\log$-enhancement.


\subsubsection{$\nu=\frac{7}{2}$ }

In the case $\nu=\frac{7}{2}$, the background fluid is described asymptotically by $w_B = -\frac{5}{9}$ when $\xi=0$. Now the slowest-dying contribution to (\ref{rho_br}) comes from the term $n=2$ in the sum. The coefficients have a different form if there is a cut-off, according to (\ref{gamman1}), however similarly as above one sees that now $w_{br} \rightarrow -1/3$ since
\ba
\rho_{br} &  \rightarrow & \frac{25H_0^{12}\eta^6}{708588\pi^2\ga_K^2}\lb 1 + \theta(K)\lp 1- 16\lp 4+\frac{\ga_K^2}{K}\rp^{-\frac{3}{2}}\rp \rb + \mathcal{O}(\eta^8)\,. \label{72c}
\ea
It is clear that the magnitude of this term depends on the infrared regulating term $\ga_K$ even when the effect of $\ga_K$ upon the background has been completely washed away.
In a closed universe, the effective backreaction density (\ref{72c}) is always positive, whereas it is proportional to the sign of $\ga^2_K$ in open and flat cases.
\newline
\newline
With a nonminimal coupling the leading order term is
\ba
\Omega_{br} &=& -\frac{G_N\lb-(1-\epsilon)H_0 \eta\rb^{\frac{2}{1-\epsilon}}35(2-3\epsilon)}{2\pi\ga_K^4\eta^6}\lb 1 + \mathcal{O}\lp(\ga_K\eta)^2\rp\rb\,, \text{when}\, K \le 0\,, \nonumber \\
\Omega_{br} &=& -\frac{G_N\lb-(1-\epsilon)H_0 \eta\rb^{\frac{2}{1-\epsilon}}210(2-3\epsilon)}{\pi\ga_K^4\eta^6}\lb 1 + \lp 8 + 5\frac{\ga_K^2}{K}\rp\lp 4+\frac{\ga_K^2}{K}\rp^{-\frac{5}{2}} + \mathcal{O}\lp(\ga_K\eta)^2\rp\rb\,, \text{when}\, K > 0\,.
\ea
The contribution is analogous to the previous case, and also now turns out to be negative for any curvature $K$ and parameter $\ga$.

\subsubsection{$\nu = \frac{9}{2}$}

When $\nu=\frac{9}{2}$, $w_B \rightarrow -\frac{5}{9}$ if $\xi=0$. Analogously to the above case, we obtain now from the $n=3$ term in the sum
\ba
\rho_{br} &  \rightarrow & \frac{735H_0^{16}\eta^8}{8589934592\pi^2\ga_K^4} + \mathcal{O}(\eta^{10} )\,, \quad K \le 0\,; \nonumber \\
\rho_{br} &  \rightarrow &  -\frac{2205H_0^{16}\eta^8}{4294967296\pi^2\ga_K^4}\lb 1 - 4\lp 8+5\frac{\ga_K^2}{K} \rp \lp 4+\frac{\ga_K^2}{K} \rp^{-\frac{5}{2}}\rb + \mathcal{O}(\eta^{10} )\,, \quad K>0\,.
\ea
Now instead the open and flat models have always positive energy density, whereas the sign depends on the ratio $\ga^2/K$ in closed models.
The next cases follow the analogous pattern: with increasing $\nu$ the magnitude of the effect gets smaller and decays more rapidly with the expansion. Again, if $\xi \neq 0$ also the $n=4$ contributes and will then dominate at late times. Given $\xi<0$, $|\Omega_{br}|$ will increase with the scale factor, but its asymptotic sign is always negative.

\subsection{Backreaction across bounces}
\label{br_bounce}

\begin{figure}
\includegraphics[width=8.75cm]{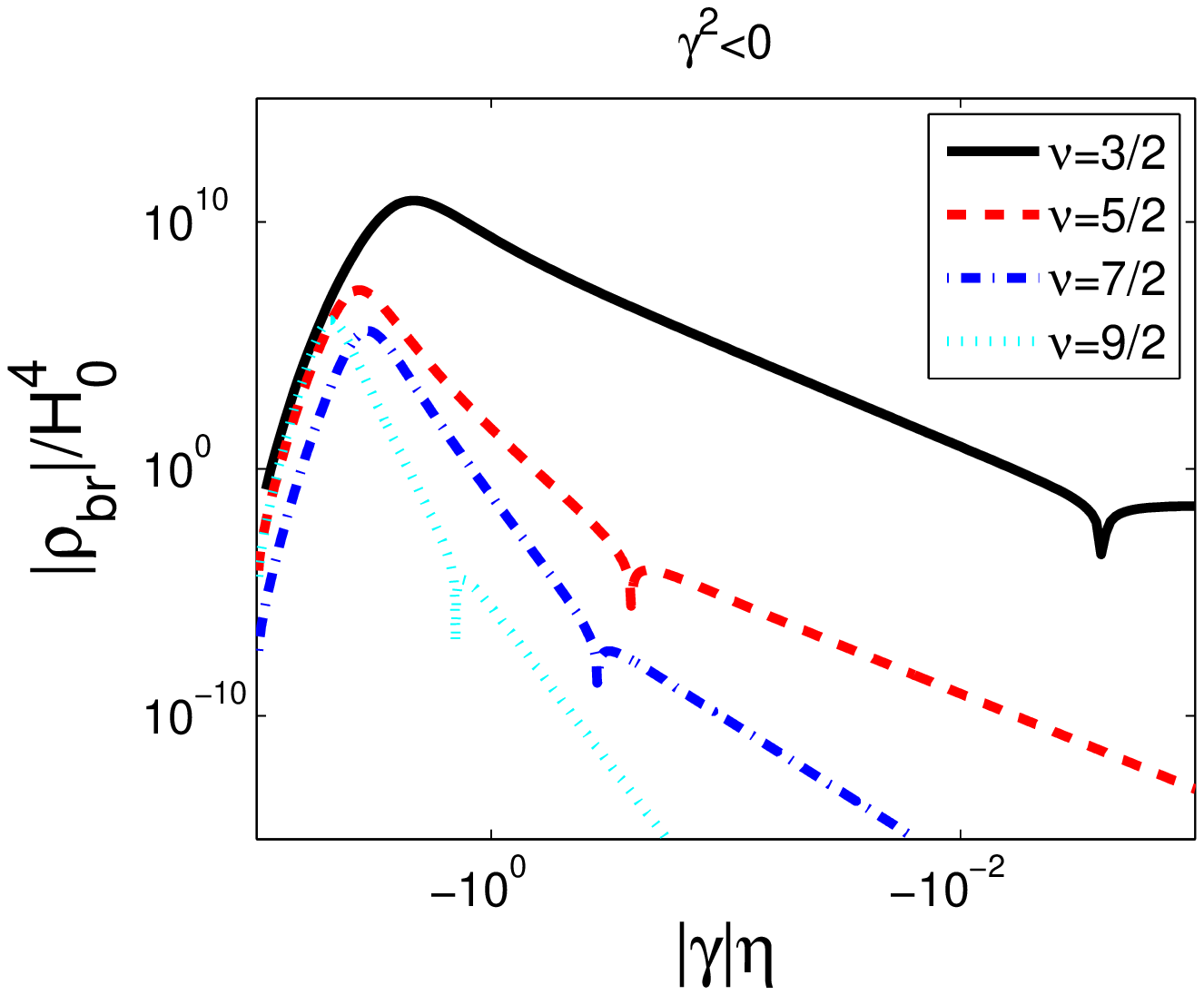}
\includegraphics[width=8.75cm]{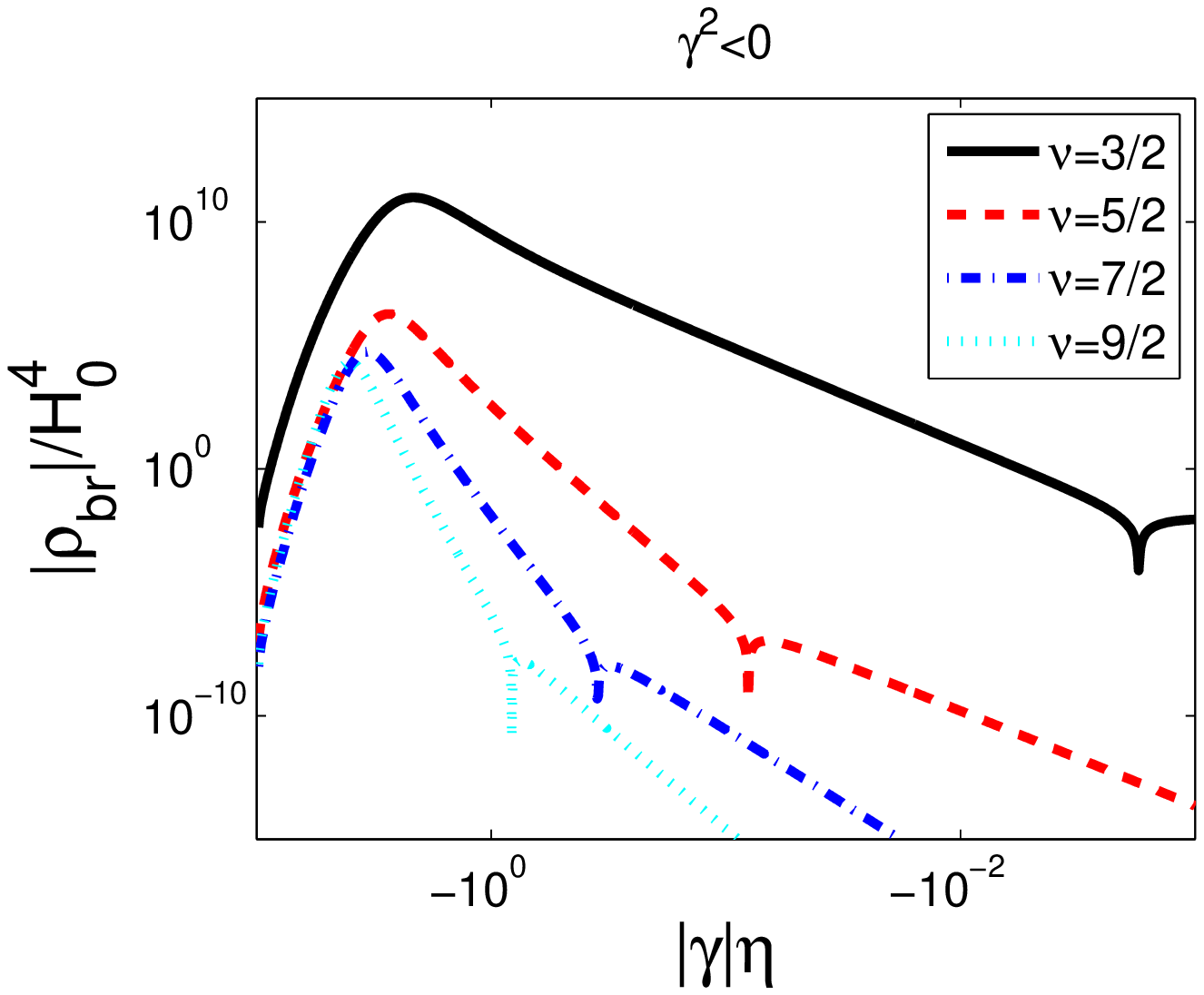}
\caption{\label{bounce_br}
Absolute value of the induced density $|\rho_{br}|$ in units of $H_0$ as a function of the conformal time in the bouncing models of the left panel in figure \ref{eps2}, with both positive and negative curvature.
{\bf Left}: Closed spatial sections, $K=-\ga^2$.
{\bf Right}: Open spatial sections, $K=\ga^2$.
}
\end{figure}
As the final application, let us briefly consider the nonsingular background evolutions described in section \ref{bounce}. Since these evolutions were constructed in such a way that they lead to inflation, the asymptotic effects at $\eta \rightarrow 0$ will be just as detailed above.
However, it is of some interest to monitor the evolution of the backreaction also at the contracting phase and across the bounce.
\newline
\newline
During contraction $\h^2 \sim K$ is a constant since $a = c_1 e^{|\ga\eta|}$. If one winds back towards the past $\eta \rightarrow -\infty$, where the size of the universe approaches infinity, it is easy to show by direct integration of the dominating terms in (\ref{tback}) that there the backreaction density scales as $\rho_{br} \sim a^{-4}$  (with a magnitude depending on $\al^{inf}$, $c_1$, $\ga$ and $K$). Thus in the asymptotic past the quantum effects are negligible. This also means that, as we approach the bounce, the backreaction density will grow. The turning point occurs at some finite $\eta$, and the relative magnitude of the various terms in (\ref{tback}) depends then on the parameters. In particular,  $\rho_{br}$ can be both positive or negative.
At the turnover $\Omega_{br}$ of course diverges. As ordinary quantum fields can acquire an effectively negative energy densities that inevitably become dynamically significant as the universe contracts, no new physics may be needed to obtain a bounce even in a flat universe.
\newline
\newline
In figure \ref{bounce_br}. we plot the backreaction energy density in the models that were shown in figure \ref{eps2}.
The fluctuations do not feature any pathologies, as $\h \rightarrow 0$.  In the left panel of the figure \ref{bounce_br}. we have negative spatial curvature, and in the right panel positive. The evolution is very similar in both cases and confirms the behaviors deduced above.
In both cases, the backreaction density grows as radiation when the universe is contracting and symmetrically decays in the expanding phase. The future asymptotic behavior is as discussed in the previous subsection. It remains to be studied in detail whether the violation of the null energy condition at the bounce could be naturally caused by the backreaction term that increases during the contracting phase, as the results here seem to suggest.



\section{Conclusions}
\label{conc_sec}

Using the operator formalism, we derived the leading quantum corrections from a non-minimally coupled massless scalar field in $D$-dimensional FLRW background. This well understood framework \cite{bd} was for the first time applied to realistic cosmological models where the spatial sections need not be exactly flat, and furthermore the universe can evolve from one phase to another. The propagator was constructed by integrating over the Fourier mode sum (or generalized Fourier modes in curved models). The result (\ref{series1}), which in most cases reduces to the simpler form (\ref{series}), was applied to derive the expectation value of the trace of the renormalized stress energy tensor for the quantum fluctuations (\ref{tback}). Also, the asymptotic energy density associated with these fluctuations (\ref{rho_br}) was computed to analyze their relevance in particular to inflationary cosmologies.
\newline
\newline
A special care was taken to obtain infrared finite results by physical means instead of automatic subtraction at play in dimensional regularization (which we employed to get ultraviolet finite answers). We found that a simple way to obtain meaningful results is to allow some spatial curvature to exist. A special example of our results is that the scalar field in a FLRW universe even with constant deceleration becomes regular in the presence of any nonzero curvature (this may also occur for large enough nonminimal coupling, see Figure 1. of Ref. \cite{Janssen:2009nz}). In general, both positive and negative spatial curvature can regulate the infrared. A negative $K$ modifies the evolution of the fluctuation modes, in a similar way as a fluid responsible for an early decelerating expansion history, and can thus render these modes less singular at very large scales. On the other hand, with a positive $K$ the very large scale modes are absent, because the eigenmodes of fluctuations are discretized in a closed universe and the perturbation wavelengths cannot be arbitrarily large.
\newline
\newline
A parameterization was introduced which allows to study analytically a wide variety models with contracting, expanding or bouncing scale factors evolutions. In particular, the parameterization describes realistic inflationary expansion histories where inflation was reached from a contracting, decelerating or curvature-dominated era.
It was found also that by taking into account such a preceding era eliminates the inflationary infrared divergences.
Then one does not have to set the modes into a Bunch-Davies vacuum in an accelerating stage in the asymptotic past, from where any physical trajectory could not have reached the observer, due to the fact that past-eternally inflating spacetimes are not geodesically complete \cite{Borde:2001nh}. Concerning specifically bouncing scenarios, our analysis shows explicitly that by assuming a bouncing mechanism, one may not avoid only the initial singularity and geodesic incompleteness associated with the background but also the infinities appearing in the fluctuation spectra (we also found that the regular backreaction density can grow and have a negative sign during the contracting phase, thus perhaps providing the mechanism for a bounce).
\newline
\newline
In this set-up then the results have straightforward physical interpretation and furthermore are not affected by any ad-hoc regularization that has to be invoked in too simplified cosmological models. We computed the leading order backreaction focusing on inflationary cosmologies. In the minimally coupled models, the backreaction tends to scale like radiation in an early expanding phase, and then scale like curvature during the inflationary phase. Exceptions are the asymptotic de Sitter case, where the backreaction is enhanced logarithmically with the scale factor, and super-inflation, where the backreaction can be power-law enhanced. The details of renormalization affect only the transient dynamics.
There sign of the induced backreaction density depends also on the background parameters, in particular the spatial curvature. However, in the cases we found the increasing backreaction during inflation, its effect was always to slow down the expansion.
In the non-minimally coupled case, a generic result is that a negative coupling parameter sets the backreaction density growing (relative to the classical background density). The main effect of non-minimal coupling is that the field ''sees'' the background expansion differently, through (\ref{xi_scale}), otherwise the coupling modifies only some finite constants in the results.

\acknowledgments

The authors thank Thomas Janssen for early involvement in the paper.
T.K. is supported by FOM and the Academy of Finland.


\appendix

\section{Field equations for $D$-dimensional FLRW metric}
\label{field_app}

The line element on a curved $D-1$-dimensional sphere can be written as
\be \label{angular}
d\Omega_{D-1}=h_{ij}d\psi^id\psi^j\,,
\ee
where $\psi_i \in [0,\pi]$ when $i=1,\dots,D-2$, and $\psi_{D-1} \in [0,2\pi)$, and the metric is given by
\be \label{angular_m}
h_{ij}=\frac{1}{K}\text{diag}\lp 1, \sin^2(\psi_1),\sin^2(\psi_1)\sin^2(\psi_2),\dots,\sin^2(\psi_1)\sin^2(\psi_2)\dots\sin^2(\psi_{D-2})\rp\,.
\ee
The role of curvature becomes more obvious when we transform the first angular coordinate into the radial coordinate $r$. These are related by $\sin(\psi_1)=\sqrt{K}r$. Using this relation, we note that
the metric in the coordinates (\ref{metric}) is generalized to $D$ dimensions straightforwardly as
\be \label{metric2}
g_{\mu\nu}dx^\mu dx^\nu = a^2(\eta)\lp -d\eta^2 +  \frac{dr^2}{1-Kr^2} + r^2d\Omega_{D-2} \rp\,,
\ee
with $\Omega_{D-2}$ given by (\ref{angular}).
In order to determine the Friedmann equations in the curved FLRW geometry, as usually we first calculate the nonvanishing Levi-Civita connection coefficients,
\ba
\Ga^{0}_{00} & = & \h\,, \qquad \Ga^{j}_{i0}=\Ga^j_{0i}=\h\delta^i_j\,, \qquad \Ga^0_{ij} = \frac{\h}{K}\delta_{ij}\sin^2(\psi_1)\dots\sin^2(\psi_{i-1})\,, \nonumber \\
\Ga^i_{ij}   & = & \Ga^i_{ji} = \left\{
\begin{array}{rl}
\cot{(\psi_j)}  & \text{if } j \le i-1\\
0 & \text{if } j  \ge i
\end{array} \right.\,,
\qquad
\Ga^j_{ii}   = \left\{
\begin{array}{rl}
-\cot{(\psi_j)}\sin^2{(\psi_j)}\dots\sin^2{(\psi_{i-1})}  & \text{if } j \le i-1\\
0 & \text{if } j  \ge i
\end{array} \right.\,.
\ea
No sum is indicated over repeated indices above. These coefficients allow us to determine the elements of the Ricci tensor and the Ricci scalar
\ba
a^2R_{00} & = & (D-1)\h'g_{00}\,; \qquad R_{0i}=0\,, \qquad a^2R_{ij} = \lb (D-2)\lp\h^2+K\rp+\h'\rb g_{ij}\,. \label{riccitensor} \\
a^2R      & = & (D-1)\lb (D-2)\lp\h^2+K\rp + 2\h' \rb\,. \label{ricciscalar}
\ea
The elements of the Einstein tensor follow immediately,
\ba
a^2G_{00} & =& -\frac{1}{2}(D-1)(D-2)\lp \h^2 + K \rp g_{00}\,, \qquad G_{0i}=0\,, \nonumber \\
a^2G_{ij} & =&  -\frac{1}{2}(D-2)\lb (D-3)\lp \h^2+ K\rp + 2\h' \rb g_{ij}\,.
\ea
We shall assume that the background stress energy tensor is an ideal fluid and thus can be written as $T^B_{\mu\nu}=u_\mu u_\nu(\rho_B+p_B)+g_{\mu\nu}p_B$, which in the fluid rest frame, specified by $u_\mu=(-a,0,0,\dots,0)$ reads just
\be
T^B_{00} = -\rho_Bg_{00}\,, \qquad T^B_{0i} = 0\,, \qquad T^B_{ij} = -p_Bg_{ij}\,.
\ee
The divergence of the stress energy tensor vanishes, implying the continuity equation
\be \label{continuityB}
\rho'_B+(D-1)\lp\rho_B+p_B\rp\h=0\,.
\ee
It is now straightforward to write the background field equations. The (00) and (ij) components imply the Friedmann equations
\be \label{friedmann}
\h^2+K = \frac{16\pi G_N a^2\rho_B}{(D-1)(D-2)}\,, \qquad \frac{1}{2}(D-3)\lp\h^2+K\rp +  \h'= -\frac{8\pi G_N a^2p_B}{(D-2)}\,,
\ee
where $G_N$ is the Newton's constant. The condition for acceleration is that $\h'$ is positive, where we can easily solve from the above pair of equations that
\be \label{doth}
\h'= - \frac{8\pi G_N a^2\left[ (D-3)\rho_B + (D-1)p_B\right]}{(D-1)(D-2)}\,.
\ee
From this we see that, just as in four spacetime dimensions, the Universe's acceleration is independent of the curvature. Furthermore, to obtain acceleration we require a sufficiently negative equation of state for the background fluid $w_B=p_B/\rho_B<-(D-3)/(D-1)$, which violates the strong energy condition that stipulates $w>-1/(D-1)$.
Next we use Eqs.(\ref{friedmann}) and (\ref{doth}) to rewrite Eq. (\ref{ricciscalar}) in terms of the background fluid as
\be \label{last}
R = \frac{16\pi G_N}{D-2}\lp 1-\lp D-1 \rp  w_B \rp\rho_B\,.
\ee
Finally we note that though we assumed a positive spatial curvature in (\ref{angular}), the formulae (\ref{riccitensor})-(\ref{last}) are valid for arbitrary $K$.

\section{Eigenfunctions of the Laplacian on a $(D-1)$-sphere}
\label{k_app}

The normal modes of fluctuations in FLRW spaces were discussed in detail in e.g. \cite{Harrison:1967zz}. Here we sketch the derivation in higher dimensions in order to make our conventions explicit and to clarify the interpretation of the wavevector $k$ through which we sum/integrate to obtain our real-space propagators. Anticipating possible future extensions of the computations at hand to higher loops, we also derive explicitly the radial parts of the eigenfunctions. These will be needed in evaluating the propagators at off-coincidence. The angular parts are independent of $K$ (and their square in the mode sum may be integrated over using 8.411.7 in \cite{gr}).

\subsection{Closed universe}

We consider the $D-1$-dimensional sphere (\ref{angular}) with the metric given by (\ref{angular_m}).
We have first assumed that the spatial curvature is positive.
The Laplacian operator derived from this metric is
\be \label{laplacian}
\nabla^2_{S^{D-1}}=K\sum_{n=1}^{D-1}\frac{\partial_{\psi_n}\sin^{D-1-n}(\psi_n)\partial_{\psi_n}}{\sin^{D-1-n}(\psi_n)\prod_{m=1}^{n-1}\sin^2(\psi_m)}\,.
\ee
It is known that on a flat sphere the eigenfunctions of the Laplacian are (higher dimensional generalizations of) spherical harmonics, 
\be \label{hh1}
\nabla^2_{S^{D-1}}\mathcal{Y}^M_\ell(\psi_1,\dots,\psi_{D-1}) = -\ell(\ell+D-2) \mathcal{Y}^M_\ell(\psi_1,\dots,\psi_{D-1})\,.
\ee
Here $M$ is a collective index for the multipole modes corresponding to $D-3$  polar and one azimuthal angle.
\newline
\newline
We want to find the eigenfunctions $Q({\bf k},{\bf x})$ of the Laplacian (\ref{laplacian}), i.e. the solutions to the Helmholz equation
\be \label{hh2}
(\nabla^2_{S^{D-1}} + \tilde{k}^2)Q({\bf k},{\bf x})=0\,.
\ee
Since merely the radial dependence changes with nonzero curvature in the metric (\ref{metric2}), the general form of the solutions will be
\be \label{decomposition}
Q({\bf k},{\bf x})=f^\ell_k(r)\mathcal{Y}^M_\ell(\psi_2,\dots,\psi_{D-1})\,,
\ee
where the label $k$ will depend on the curvature. The convenient choice will turn out to be
\be \label{ktilde}
k^2=\tilde{k}^2+\lp\frac{D-2}{2}\rp^2K\,.
\ee
The Laplacian decomposes into
\be
\nabla^2_{S^{D-1}} = \frac{\sqrt{1-Kr^2}}{(\sqrt{K}r)^{D-2}}\partial_r\lp (\sqrt{K}r)^{D-2} \sqrt{1-Kr^2} \rp \partial_r + \frac{1}{r^2}\nabla^2_{S^{D-2}}\,.
\ee
We are then ready to write down the equation for the radial part of the eigenfunction. In terms of the $\psi_1$, it becomes
\be
\lb\frac{1}{\sin^{D-2}(\psi_1)}\partial_{\psi_1} \sin^{D-2}(\psi_1)\partial_{\psi_1} + \frac{\tilde{k}^2}{K} + \frac{\ell(\ell+D-3)}{\sin^2(\psi_1)}\rb f^\ell_k(\psi_1) = 0\,.
\ee
The solutions to this equation can be written in terms of the associated Legendre functions,
\be
f^\ell_k(\psi_1) = \sin^{-\frac{D-3}{2}}(\psi_1)\lb c_1 P^\mu_\la \lp\cos(\psi_1)\rp + c_2 Q^\mu_\la \lp\cos(\psi_1)\rp \rb\,,
\ee
where
\ba
\la & = & -\frac{1}{2}+\sqrt{\frac{\tilde{k}^2}{K}+\lp\frac{D-2}{2}\rp^2} = -\frac{1}{2}+\frac{k}{\sqrt{K}}\,, \\
\mu & = & \ell+\frac{1}{2}(D-3) \,.
\ea
At least at $D=4$, it occurs that $\mu$ is a half integer, and we can use, instead of $P^\mu_\la$ and $Q^\mu_\la$, the pair $P^{\pm\mu}_\la$ as independent solutions, see formula
see 8.737.1 in \cite{gr}. By requiring regularity at the origin, using formula 8.756.1 of \cite{gr}, we are left with only $P^{-\mu}_\la$. Thus
\be
f^\ell_k(\psi_1) = N^\ell_k\sin^{-\frac{D-3}{2}}(\psi_1) P^{-\mu}_\la \lp\cos(\psi_1)\rp,
\ee
where $N^\ell_k$ is a normalization constant.
The normalization may be fixed as
\be
N^\ell_k = \sqrt{\frac{(2k+1)\Gamma(k/\sqrt{K}-\mu + \frac{1}{2})}{2\Gamma(k/\sqrt{K}+\mu + \frac{1}{2})}}\,,
\ee
corresponding to the measure
\be
\int_{-1}^{+1}dx\sin^{D-3}(x) f^\ell_k(x)f^{\ell'}_{k'}(x) = \delta_{k,k'}\delta_{\ell,\ell'}\,.
\ee
Periodicity requires that $k/\sqrt{K}$ is an integer in $D=4$, and its lowest eigenmode is $k=D\sqrt{K}/2$. This can be seen as follows. The eigenfunction $f^\ell_k(\psi_1)$ is single-valued
when (see 8.737.2 in \cite{gr}),
\be
f^\ell_k(-\psi_1)=\cos{\lb(\la-\mu)\pi\rb}f^\ell_k(\psi_1)\,.
\ee
It follows that $\tilde{k}^2/K=n(n+1)$, where $n$ is a positive integer. The fundamental mode is $n=1$, and the values of $k$ and $\ell$ are then discretized as
\be
\frac{k}{\sqrt{K}}=n+\frac{1}{2}(D-2)\,, \quad -n \le \ell \le n\,, \quad n=1,2,\dots\,.
\ee
A function may then be expressed as
\be
F(\eta,\psi_1,\dots,\psi_{D-1})= \sum_{n=1}^{\infty}\sum_{\ell,M} F(\eta,(n+\frac{1}{2}(D-2))\sqrt{K})f^\ell_{(n+\frac{D-2}{2})\sqrt{K}}(\psi_1)a_M(\eta)\mathcal{Y}^M_\ell(\psi_2,\dots,\psi_{D-1})\,.
\ee
Replacing the sum with by integral is an approximation that becomes increasingly good in the UV. We may then write the Fourier transformation (a bit schematically) as
\be
F(\eta,{\bf x})= \frac{1}{(2\pi)^\frac{D-1}{2}}\int_{\frac{D}{2}\sqrt{K}}^{\infty}d^{D-1} k F(\eta,{\bf k})Q({\bf k},{\bf x})\,.
\ee
In terms of the wavemode $k$, shifted with respect to the eigenmode appearing in (\ref{hh1}), the full d'Alembertian acting on a scalar function $F$ gives yields
\be
\Box F = -\frac{1}{a^2}\lp F''+(D-2)\h F'+ k^2 F - \frac{1}{4}(D-2)^2KF\rp\,, \label{dalembertian}
\ee
leading to our result (\ref{eom}) in the case of non-minimally coupled scalar field.

\subsection{Open universe, $K<0$}

With negative curvature, the analysis is quite analogous. The results can be read from above by analytic continuation.
In particular, the mode functions become then
\be
f^\ell_k(\psi_1) = \sqrt{\frac{(2k+1)\Gamma(k/\sqrt{-K}-\frac{1}{2}(D-2)-\ell)}{2\Gamma(k/\sqrt{-K}+ \frac{1}{2}(D-4)+\ell)}}  \sinh^{-\frac{D-3}{2}}(\psi_1) P_{\frac{1}{2}-\frac{k}{\sqrt{-K}}}^{-\frac{1}{2}(D-3)-\ell} \lp\cosh(\psi_1)\rp\,,
\ee
and thus
\be
\int_{-1}^{+1}dx\sinh^{D-3}(x) f^\ell_k(x)f^{\ell'}_{k'}(x) = \delta(k-k')\delta_{\ell,\ell'}\,.
\ee
Now the spectrum is continuous as there are no periodicity conditions. Now $\tilde{k}>(D-2)\sqrt{-K}/2$, thus $k>0$\footnote{Though the basis is complete for these values, and larger modes are not generated by a scalar field, it is interesting to note that the most general homogeneous Gaussian random field involves the modes up from $\tilde{k}>0$. This has been clarified in \cite{Lyth:1995cw}.}. The upshot is that by using the index $k$, we can consider all positive modes as relevant and don't need to introduce a nonzero IR cut-off. When matching a particular wave-mode to a physical observable, the subtle question would arise, which is the effective physical wavelength of a perturbation, $2\pi/k$ or $2\pi/\tilde{k}$? However, in the present study this ambiguity does not arise as we are considering real-space backreaction quantities, integrated over all Fourier modes.

\section{The propagator when $K \le 0$ or $\ga^2>0$}
\label{special}

Here we consider the propagator, barring the case that $K>0>\ga_K^2$. Equation (\ref{series}) can then be written as
\be
i\Delta(x;x)  =  \Delta_\nu - \theta(K)i\delta\Delta_\nu\,,
\ee
where
\be
i\Delta_\nu =  \frac{\ga_K^{D-2}}{a^{D-2}2^D \pi^{\frac{D}{2}}}\sum_{n=0}^{\nu-\frac{1}{2}}\frac{\Ga(1-\frac{D}{2}+n)\Ga(\nu+\frac{1}{2}+n)}{n!\Ga(\nu+\frac{1}{2}-n)}\frac{1}{(\ga_K\eta)^{2n}}\,,
\ee
and $\delta\Delta_\nu$ appears in closed universes $K>0$ to subtract the supercurvature modes.
The general expression for it can be written in terms of the incomplete beta functions
\be \label{deltadelta}
i\delta\Delta_\nu = \frac{(-1)^{\frac{D-1}{2}}\ga_K^{D-2}}{a^{D-2}2^D\pi^{\frac{D}{2}}\Ga(\frac{D-1}{2})}\sum_{n}^{\nu-\frac{1}{2}}
\frac{\Ga(\frac{1}{2}+n)\Ga(\nu+\frac{1}{2}+n)}{n!\Ga(\nu+\frac{1}{2}-n)}
B_{\frac{-D^2K}{4\ga_K^2}}\lp\frac{1}{2}(D-1),\frac{1}{2}-n\rp \frac{1}{(\ga_K\eta)^{2n}}\,.
\ee
In the following we evaluate exactly the propagator at a few special values of $\nu$:
\be
i\Delta_\frac{3}{2} =
\frac{\Gamma \left(1-\frac{D}{2}\right)}{(2a)^{D-2}\pi^{D}\ga_K^{5-D}\eta^2}\left(\ga_K^2 \eta^2-D+2\right) \,,
\ee
\be
i\Delta_\frac{5}{2} =
\frac{\Gamma \left(1-\frac{D}{2}\right)}{(2a)^{D-2}\pi^{D}\ga_K^{7-D}\eta^4}
\left(\ga_K^4 \eta^4-3 (D-2)\ga_K^2 \eta^2+3 (D-4) (D-2)\right)\,,
\ee
\be
i\Delta_\frac{7}{2} =
\frac{\Gamma \left(1-\frac{D}{2}\right)}{(2a)^{D-2}\pi^{D}\ga_K^{9-D}\eta^6}
  \left(\ga_K^6 \eta^6-6 (D-2)\ga_K^4 \eta^4+15 (D-4) (D-2)\ga_K^2 \eta^2-15 (D-6) (D-4) (D-2)\right) \,.
\ee
\newline
\newline
Then consider the terms which take into account the cut-off.
In $D=4$, Eq. (\ref{deltadelta}) becomes as follows
\be
i\delta\Delta_\nu =
\frac{K^{\frac{3}{2}}} {64\pi ^2\ga_K  a^4 } \sum _{n=0}^{\nu -\frac{1}{2}} \frac{ \Gamma \left(\frac{1}{2}+n\right) \Gamma \left(\nu +\frac{1}{2}+n\right) }{n! \Gamma \left(\nu +\frac{1}{2}-n\right)}\,_2F_1\left(\frac{3}{2},\frac{1}{2}+n;\frac{5}{2};-\frac{4 K}{\ga_K^2}\right) \frac{1}{(\ga_K \eta)^{2n}}\,.
\ee
In the case $\nu=\frac{3}{2}$ this reads
\be \label{32del}
i\delta\Delta_\frac{3}{2} =
\frac{6 \ga_K \sqrt{K} \sqrt{\frac{4 K}{\ga_K^2}+1} \left(\left(\ga_K^2+4 K\right) \eta^2-2\right)-3 \left(\ga_K^2+4 K\right) \left((\ga_K\eta)^2-2\right) \sinh ^{-1}\left(\frac{2
   \sqrt{K}}{\ga_K}\right)}{1024  \left(\ga_K^2+4 K\right) \pi ^{\frac{3}{2}}a^4 \eta^2}
\ee
The next couple of cases are
\be
i\delta\Delta_\frac{5}{2} =
\frac{6 \sqrt{K} \left(\left(\ga_K^3+4 K\ga_K\right)^2 \eta^4-6 \ga_K^2 \left(\ga_K^2+4 K\right) \eta^2+24 K\right)-3 \ga_K^3 \left(\ga_K^2+4 K\right) \sqrt{\frac{4 K}{\ga_K^2}+1} \eta^2
   \left(\ga_K^2 \eta^2-6\right) \sinh ^{-1}\left(\frac{2 \sqrt{K}}{\ga_K}\right)}{1024  \ga_K^3 \left(\ga_K^2+4 K\right) \sqrt{\frac{4 K}{\ga_K^2}+1} \pi^\frac{3}{2}( a\eta)^4}\,,
\ee
\ba
i\delta\Delta_\frac{7}{2} & = &
\frac{1}{1024 \left(\ga_K^3+4 K \ga_K\right)^3 \pi
   ^{\frac{3}{2}}  a^4 \eta^6}
\Big[6 \sqrt{K} \sqrt{\frac{4 K}{\ga_K^2}+1} \Big(\ga_K^4 \left(\ga_K^2+4 K\right)^3 \eta^6-12 \ga_K^4 \left(\ga_K^2+4 K\right)^2 \eta^4 \nonumber \\
& +& 120 \ga_K^2 K \left(\ga_K^2+4 K\right) \eta^2+120 K \left(5
   \ga_K^2+8 K\right)\Big)-3 \left(\ga_K^3+4 K \ga_K\right)^3 \eta^4 \left(\ga_K^2 \eta^2-12\right) \sinh ^{-1}\left(\frac{2 \sqrt{K}}{\ga_K}\right)\Big]\,.
\ea
These special cases are analyzed and evaluated numerically in section \ref{ba_sec}.

\bibliography{epsilon}

\begin{thebibliography}{10}%
\makeatletter
\providecommand \@ifxundefined [1]{%
 \ifx #1\undefined \expandafter \@firstoftwo
 \else \expandafter \@secondoftwo
\fi
}%
\providecommand \@ifnum [1]{%
 \ifnum #1\expandafter \@firstoftwo
 \else \expandafter \@secondoftwo
\fi
}%
\providecommand \enquote [1]{``#1''}%
\providecommand \bibnamefont  [1]{#1}%
\providecommand \bibfnamefont [1]{#1}%
\providecommand \citenamefont [1]{#1}%
\providecommand\href[0]{\@sanitize\@href}%
\providecommand\@href[1]{\endgroup\@@startlink{#1}\endgroup\@@href}%
\providecommand\@@href[1]{#1\@@endlink}%
\providecommand \@sanitize [0]{\begingroup\catcode`\&12\catcode`\#12\relax}%
\@ifxundefined \pdfoutput {\@firstoftwo}{%
 \@ifnum{\z@=\pdfoutput}{\@firstoftwo}{\@secondoftwo}%
}{%
 \providecommand\@@startlink[1]{\leavevmode\special{html:<a href="#1">}}%
 \providecommand\@@endlink[0]{\special{html:</a>}}%
}{%
 \providecommand\@@startlink[1]{%
  \leavevmode
  \pdfstartlink
   attr{/Border[0 0 1 ]/H/I/C[0 1 1]}%
   user{/Subtype/Link/A<</Type/Action/S/URI/URI(#1)>>}%
  \relax
 }%
 \providecommand\@@endlink[0]{\pdfendlink}%
}%
\providecommand \url  [0]{\begingroup\@sanitize \@url }%
\providecommand \@url [1]{\endgroup\@href {#1}{\urlprefix}}%
\providecommand \urlprefix [0]{URL }%
\providecommand \Eprint[0]{\href }%
\@ifxundefined \urlstyle {%
  \providecommand \doi [1]{doi:\discretionary{}{}{}#1}%
}{%
  \providecommand \doi [0]{doi:\discretionary{}{}{}\begingroup
  \urlstyle{rm}\Url }%
}%
\providecommand \doibase [0]{http://dx.doi.org/}%
\providecommand \Doi[1]{\href{\doibase#1}}%
\providecommand \bibAnnote [3]{%
  \BibitemShut{#1}%
  \begin{quotation}\noindent
    \textsc{Key:}\ #2\\\textsc{Annotation:}\ #3%
  \end{quotation}%
}%
\providecommand \bibAnnoteFile [2]{%
  \IfFileExists{#2}{\bibAnnote {#1} {#2} {\input{#2}}}{}%
}%
\providecommand \typeout [0]{\immediate \write \m@ne }%
\providecommand \selectlanguage [0]{\@gobble}%
\providecommand \bibinfo [0]{\@secondoftwo}%
\providecommand \bibfield [0]{\@secondoftwo}%
\providecommand \translation [1]{[#1]}%
\providecommand \BibitemOpen[0]{}%
\providecommand \bibitemStop [0]{}%
\providecommand \bibitemNoStop [0]{.\EOS\space}%
\providecommand \EOS [0]{\spacefactor3000\relax}%
\providecommand \BibitemShut [1]{\csname bibitem#1\endcsname}%
\bibitem{Komatsu:2008hk}%
  \BibitemOpen
  \bibfield{author}{%
  \bibinfo {author} {\bibfnamefont{E.}~\bibnamefont{Komatsu}} \emph{et~al.}
  (\bibinfo {collaboration} {WMAP}),\ }%
  \bibfield{journal}{%
  \Doi{10.1088/0067-0049/180/2/330}{\bibinfo {journal} {Astrophys. J. Suppl.}}\
  }%
  \textbf{\bibinfo {volume} {180}},\ \bibinfo {pages} {330} (\bibinfo {year}
  {2009}),\ \Eprint{http://arxiv.org/abs/0803.0547}{arXiv:0803.0547
  [astro-ph]}%
  \bibAnnoteFile{NoStop}{Komatsu:2008hk}%
\bibitem{Guth:1980zm}%
  \BibitemOpen
  \bibfield{author}{%
  \bibinfo {author} {\bibfnamefont{A.~H.}\ \bibnamefont{Guth}},\ }%
  \bibfield{journal}{%
  \Doi{10.1103/PhysRevD.23.347}{\bibinfo {journal} {Phys. Rev.}}\ }%
  \textbf{\bibinfo {volume} {D23}},\ \bibinfo {pages} {347} (\bibinfo {year}
  {1981})%
  \bibAnnoteFile{NoStop}{Guth:1980zm}%
\bibitem{Starobinsky:1982ee}%
  \BibitemOpen
  \bibfield{author}{%
  \bibinfo {author} {\bibfnamefont{A.~A.}\ \bibnamefont{Starobinsky}},\ }%
  \bibfield{journal}{%
  \Doi{10.1016/0370-2693(82)90541-X}{\bibinfo {journal} {Phys. Lett.}}\ }%
  \textbf{\bibinfo {volume} {B117}},\ \bibinfo {pages} {175} (\bibinfo {year}
  {1982})%
  \bibAnnoteFile{NoStop}{Starobinsky:1982ee}%
\bibitem{Vilenkin:1982wt}%
  \BibitemOpen
  \bibfield{author}{%
  \bibinfo {author} {\bibfnamefont{A.}~\bibnamefont{Vilenkin}}\ and\ \bibinfo
  {author} {\bibfnamefont{L.~H.}\ \bibnamefont{Ford}},\ }%
  \bibfield{journal}{%
  \Doi{10.1103/PhysRevD.26.1231}{\bibinfo {journal} {Phys. Rev.}}\ }%
  \textbf{\bibinfo {volume} {D26}},\ \bibinfo {pages} {1231} (\bibinfo {year}
  {1982})%
  \bibAnnoteFile{NoStop}{Vilenkin:1982wt}%
\bibitem{Grishchuk:1974ny}%
  \BibitemOpen
  \bibfield{author}{%
  \bibinfo {author} {\bibfnamefont{L.~P.}\ \bibnamefont{Grishchuk}},\ }%
  \bibfield{journal}{%
  \bibinfo {journal} {Sov. Phys. JETP}\ }%
  \textbf{\bibinfo {volume} {40}},\ \bibinfo {pages} {409} (\bibinfo {year}
  {1975})%
  \bibAnnoteFile{NoStop}{Grishchuk:1974ny}%
\bibitem{Janssen:2008dw}%
  \BibitemOpen
  \bibfield{author}{%
  \bibinfo {author} {\bibfnamefont{T.}~\bibnamefont{Janssen}}\ and\ \bibinfo
  {author} {\bibfnamefont{T.}~\bibnamefont{Prokopec}},\ }%
  \bibfield{journal}{%
  \Doi{10.1016/j.aop.2009.09.003}{\bibinfo {journal} {Annals Phys.}}\ }%
  \textbf{\bibinfo {volume} {325}},\ \bibinfo {pages} {948} (\bibinfo {year}
  {2010}),\ \Eprint{http://arxiv.org/abs/0807.0447}{arXiv:0807.0447 [gr-qc]}%
  \bibAnnoteFile{NoStop}{Janssen:2008dw}%
\bibitem{Tsamis:2006gj}%
  \BibitemOpen
  \bibfield{author}{%
  \bibinfo {author} {\bibfnamefont{N.~C.}\ \bibnamefont{Tsamis}}\ and\ \bibinfo
  {author} {\bibfnamefont{R.~P.}\ \bibnamefont{Woodard}},\ }%
  \bibfield{journal}{%
  \Doi{10.1063/1.2738361}{\bibinfo {journal} {J. Math. Phys.}}\ }%
  \textbf{\bibinfo {volume} {48}},\ \bibinfo {pages} {052306} (\bibinfo {year}
  {2007}),\ \Eprint{http://arxiv.org/abs/gr-qc/0608069}{arXiv:gr-qc/0608069}%
  \bibAnnoteFile{NoStop}{Tsamis:2006gj}%
\bibitem{Janssen:2007yu}%
  \BibitemOpen
  \bibfield{author}{%
  \bibinfo {author} {\bibfnamefont{T.}~\bibnamefont{Janssen}}\ and\ \bibinfo
  {author} {\bibfnamefont{T.}~\bibnamefont{Prokopec}},\ }%
  \bibfield{journal}{%
  \bibinfo {journal} {JCAP}\ }%
  \textbf{\bibinfo {volume} {0705}},\ \bibinfo {pages} {010} (\bibinfo {year}
  {2007}),\ \Eprint{http://arxiv.org/abs/gr-qc/0703050}{arXiv:gr-qc/0703050}%
  \bibAnnoteFile{NoStop}{Janssen:2007yu}%
\bibitem{Germani:2009gg}%
  \BibitemOpen
  \bibfield{author}{%
  \bibinfo {author} {\bibfnamefont{C.}~\bibnamefont{Germani}}\ and\ \bibinfo
  {author} {\bibfnamefont{A.}~\bibnamefont{Kehagias}},\ }%
  \bibfield{journal}{%
  \Doi{10.1088/1475-7516/2009/11/005}{\bibinfo {journal} {JCAP}}\ }%
  \textbf{\bibinfo {volume} {0911}},\ \bibinfo {pages} {005} (\bibinfo {year}
  {2009}),\ \Eprint{http://arxiv.org/abs/0908.0001}{arXiv:0908.0001
  [astro-ph.CO]}%
  \bibAnnoteFile{NoStop}{Germani:2009gg}%
\bibitem{Koivisto:2009fb}%
  \BibitemOpen
  \bibfield{author}{%
  \bibinfo {author} {\bibfnamefont{T.~S.}\ \bibnamefont{Koivisto}}\ and\
  \bibinfo {author} {\bibfnamefont{N.~J.}\ \bibnamefont{Nunes}},\ }%
  \bibfield{journal}{%
  \Doi{10.1103/PhysRevD.80.103509}{\bibinfo {journal} {Phys. Rev.}}\ }%
  \textbf{\bibinfo {volume} {D80}},\ \bibinfo {pages} {103509} (\bibinfo {year}
  {2009}),\ \Eprint{http://arxiv.org/abs/0908.0920}{arXiv:0908.0920
  [astro-ph.CO]}%
  \bibAnnoteFile{NoStop}{Koivisto:2009fb}%
\bibitem{Tsamis:1996qm}%
  \BibitemOpen
  \bibfield{author}{%
  \bibinfo {author} {\bibfnamefont{N.~C.}\ \bibnamefont{Tsamis}}\ and\ \bibinfo
  {author} {\bibfnamefont{R.~P.}\ \bibnamefont{Woodard}},\ }%
  \bibfield{journal}{%
  \Doi{10.1006/aphy.1997.5613}{\bibinfo {journal} {Annals Phys.}}\ }%
  \textbf{\bibinfo {volume} {253}},\ \bibinfo {pages} {1} (\bibinfo {year}
  {1997}),\ \Eprint{http://arxiv.org/abs/hep-ph/9602316}{arXiv:hep-ph/9602316}%
  \bibAnnoteFile{NoStop}{Tsamis:1996qm}%
\bibitem{Mukhanov:1996ak}%
  \BibitemOpen
  \bibfield{author}{%
  \bibinfo {author} {\bibfnamefont{V.~F.}\ \bibnamefont{Mukhanov}}, \bibinfo
  {author} {\bibfnamefont{L.~R.~W.}\ \bibnamefont{Abramo}},\ and\ \bibinfo
  {author} {\bibfnamefont{R.~H.}\ \bibnamefont{Brandenberger}},\ }%
  \bibfield{journal}{%
  \Doi{10.1103/PhysRevLett.78.1624}{\bibinfo {journal} {Phys. Rev. Lett.}}\ }%
  \textbf{\bibinfo {volume} {78}},\ \bibinfo {pages} {1624} (\bibinfo {year}
  {1997}),\ \Eprint{http://arxiv.org/abs/gr-qc/9609026}{arXiv:gr-qc/9609026}%
  \bibAnnoteFile{NoStop}{Mukhanov:1996ak}%
\bibitem{Bunch:1977sq}%
  \BibitemOpen
  \bibfield{author}{%
  \bibinfo {author} {\bibfnamefont{T.~S.}\ \bibnamefont{Bunch}}\ and\ \bibinfo
  {author} {\bibfnamefont{P.~C.~W.}\ \bibnamefont{Davies}},\ }%
  \bibfield{journal}{%
  \bibinfo {journal} {Proc. Roy. Soc. Lond.}\ }%
  \textbf{\bibinfo {volume} {A357}},\ \bibinfo {pages} {381} (\bibinfo {year}
  {1977})%
  \bibAnnoteFile{NoStop}{Bunch:1977sq}%
\bibitem{bd}%
  \BibitemOpen
  \bibfield{author}{%
  \bibinfo {author} {\bibfnamefont{N.}~\bibnamefont{Birrel}}\ and\ \bibinfo
  {author} {\bibfnamefont{P.~C.~W.}\ \bibnamefont{Davies}},\ }%
  \emph{\bibinfo {title} {{Quantum Fields in Curved Space}}}\ (\bibinfo
  {publisher} {Cambridge, UK: University Press},\ \bibinfo {year} {1982})%
  \bibAnnoteFile{NoStop}{bd}%
\bibitem{Vilenkin:1983xp}%
  \BibitemOpen
  \bibfield{author}{%
  \bibinfo {author} {\bibfnamefont{A.}~\bibnamefont{Vilenkin}},\ }%
  \bibfield{journal}{%
  \Doi{10.1016/0550-3213(83)90208-0}{\bibinfo {journal} {Nucl. Phys.}}\ }%
  \textbf{\bibinfo {volume} {B226}},\ \bibinfo {pages} {527} (\bibinfo {year}
  {1983})%
  \bibAnnoteFile{NoStop}{Vilenkin:1983xp}%
\bibitem{Tsamis:1993ub}%
  \BibitemOpen
  \bibfield{author}{%
  \bibinfo {author} {\bibfnamefont{N.~C.}\ \bibnamefont{Tsamis}}\ and\ \bibinfo
  {author} {\bibfnamefont{R.~P.}\ \bibnamefont{Woodard}},\ }%
  \bibfield{journal}{%
  \Doi{10.1088/0264-9381/11/12/012}{\bibinfo {journal} {Class. Quant. Grav.}}\
  }%
  \textbf{\bibinfo {volume} {11}},\ \bibinfo {pages} {2969} (\bibinfo {year}
  {1994})%
  \bibAnnoteFile{NoStop}{Tsamis:1993ub}%
\bibitem{Linde:1982zj}%
  \BibitemOpen
  \bibfield{author}{%
  \bibinfo {author} {\bibfnamefont{A.~D.}\ \bibnamefont{Linde}},\ }%
  \bibfield{journal}{%
  \Doi{10.1016/0370-2693(82)90086-7}{\bibinfo {journal} {Phys. Lett.}}\ }%
  \textbf{\bibinfo {volume} {B114}},\ \bibinfo {pages} {431} (\bibinfo {year}
  {1982})%
  \bibAnnoteFile{NoStop}{Linde:1982zj}%
\bibitem{Losic:2006ht}%
  \BibitemOpen
  \bibfield{author}{%
  \bibinfo {author} {\bibfnamefont{B.}~\bibnamefont{Losic}}\ and\ \bibinfo
  {author} {\bibfnamefont{W.~G.}\ \bibnamefont{Unruh}},\ }%
  \bibfield{journal}{%
  \Doi{10.1103/PhysRevD.74.023511}{\bibinfo {journal} {Phys. Rev.}}\ }%
  \textbf{\bibinfo {volume} {D74}},\ \bibinfo {pages} {023511} (\bibinfo {year}
  {2006}),\ \Eprint{http://arxiv.org/abs/gr-qc/0604122}{arXiv:gr-qc/0604122}%
  \bibAnnoteFile{NoStop}{Losic:2006ht}%
\bibitem{Losic:2008ht}%
  \BibitemOpen
  \bibfield{author}{%
  \bibinfo {author} {\bibfnamefont{B.}~\bibnamefont{Losic}}\ and\ \bibinfo
  {author} {\bibfnamefont{W.~G.}\ \bibnamefont{Unruh}},\ }%
  \bibfield{journal}{%
  \Doi{10.1103/PhysRevLett.101.111101}{\bibinfo {journal} {Phys. Rev. Lett.}}\
  }%
  \textbf{\bibinfo {volume} {101}},\ \bibinfo {pages} {111101} (\bibinfo {year}
  {2008}),\ \Eprint{http://arxiv.org/abs/0804.4296}{arXiv:0804.4296 [gr-qc]}%
  \bibAnnoteFile{NoStop}{Losic:2008ht}%
\bibitem{Janssen:2007ht}%
  \BibitemOpen
  \bibfield{author}{%
  \bibinfo {author} {\bibfnamefont{T.}~\bibnamefont{Janssen}}\ and\ \bibinfo
  {author} {\bibfnamefont{T.}~\bibnamefont{Prokopec}},\ }%
  \bibfield{journal}{%
  \bibinfo {journal} {Class. Quant. Grav.}\ }%
  \textbf{\bibinfo {volume} {25}},\ \bibinfo {pages} {055007} (\bibinfo {year}
  {2008}),\ \Eprint{http://arxiv.org/abs/0707.3919}{arXiv:0707.3919 [gr-qc]}%
  \bibAnnoteFile{NoStop}{Janssen:2007ht}%
\bibitem{Janssen:2009nz}%
  \BibitemOpen
  \bibfield{author}{%
  \bibinfo {author} {\bibfnamefont{T.~M.}\ \bibnamefont{Janssen}}\ and\
  \bibinfo {author} {\bibfnamefont{T.}~\bibnamefont{Prokopec}}}%
   (\bibinfo {year} {2009}),\
  \Eprint{http://arxiv.org/abs/0906.0666}{arXiv:0906.0666 [gr-qc]}%
  \bibAnnoteFile{NoStop}{Janssen:2009nz}%
\bibitem{Janssen:2008px}%
  \BibitemOpen
  \bibfield{author}{%
  \bibinfo {author} {\bibfnamefont{T.~M.}\ \bibnamefont{Janssen}}, \bibinfo
  {author} {\bibfnamefont{S.~P.}\ \bibnamefont{Miao}}, \bibinfo {author}
  {\bibfnamefont{T.}~\bibnamefont{Prokopec}},\ and\ \bibinfo {author}
  {\bibfnamefont{R.~P.}\ \bibnamefont{Woodard}},\ }%
  \bibfield{journal}{%
  \Doi{10.1088/0264-9381/25/24/245013}{\bibinfo {journal} {Class. Quant.
  Grav.}}\ }%
  \textbf{\bibinfo {volume} {25}},\ \bibinfo {pages} {245013} (\bibinfo {year}
  {2008}),\ \Eprint{http://arxiv.org/abs/0808.2449}{arXiv:0808.2449 [gr-qc]}%
  \bibAnnoteFile{NoStop}{Janssen:2008px}%
\bibitem{Linde:1982uu}%
  \BibitemOpen
  \bibfield{author}{%
  \bibinfo {author} {\bibfnamefont{A.~D.}\ \bibnamefont{Linde}},\ }%
  \bibfield{journal}{%
  \Doi{10.1016/0370-2693(82)90293-3}{\bibinfo {journal} {Phys. Lett.}}\ }%
  \textbf{\bibinfo {volume} {B116}},\ \bibinfo {pages} {335} (\bibinfo {year}
  {1982})%
  \bibAnnoteFile{NoStop}{Linde:1982uu}%
\bibitem{Giddings:2010nc}%
  \BibitemOpen
  \bibfield{author}{%
  \bibinfo {author} {\bibfnamefont{S.~B.}\ \bibnamefont{Giddings}}\ and\
  \bibinfo {author} {\bibfnamefont{M.~S.}\ \bibnamefont{Sloth}}}%
   (\bibinfo {year} {2010}),\
  \Eprint{http://arxiv.org/abs/1005.1056}{arXiv:1005.1056 [hep-th]}%
  \bibAnnoteFile{NoStop}{Giddings:2010nc}%
\bibitem{Byrnes:2010yc}%
  \BibitemOpen
  \bibfield{author}{%
  \bibinfo {author} {\bibfnamefont{C.~T.}\ \bibnamefont{Byrnes}}, \bibinfo
  {author} {\bibfnamefont{M.}~\bibnamefont{Gerstenlauer}}, \bibinfo {author}
  {\bibfnamefont{A.}~\bibnamefont{Hebecker}}, \bibinfo {author}
  {\bibfnamefont{S.}~\bibnamefont{Nurmi}},\ and\ \bibinfo {author}
  {\bibfnamefont{G.}~\bibnamefont{Tasinato}}}%
   (\bibinfo {year} {2010}),\
  \Eprint{http://arxiv.org/abs/1005.3307}{arXiv:1005.3307 [hep-th]}%
  \bibAnnoteFile{NoStop}{Byrnes:2010yc}%
\bibitem{Senatore:2009cf}%
  \BibitemOpen
  \bibfield{author}{%
  \bibinfo {author} {\bibfnamefont{L.}~\bibnamefont{Senatore}}\ and\ \bibinfo
  {author} {\bibfnamefont{M.}~\bibnamefont{Zaldarriaga}}}%
   (\bibinfo {year} {2009}),\
  \Eprint{http://arxiv.org/abs/0912.2734}{arXiv:0912.2734 [hep-th]}%
  \bibAnnoteFile{NoStop}{Senatore:2009cf}%
\bibitem{Kahya:2010xh}%
  \BibitemOpen
  \bibfield{author}{%
  \bibinfo {author} {\bibfnamefont{E.~O.}\ \bibnamefont{Kahya}}, \bibinfo
  {author} {\bibfnamefont{V.~K.}\ \bibnamefont{Onemli}},\ and\ \bibinfo
  {author} {\bibfnamefont{R.~P.}\ \bibnamefont{Woodard}}}%
   (\bibinfo {year} {2010}),\
  \Eprint{http://arxiv.org/abs/1006.3999}{arXiv:1006.3999 [astro-ph.CO]}%
  \bibAnnoteFile{NoStop}{Kahya:2010xh}%
\bibitem{Seery:2010kh}%
  \BibitemOpen
  \bibfield{author}{%
  \bibinfo {author} {\bibfnamefont{D.}~\bibnamefont{Seery}},\ }%
  \bibfield{journal}{%
  \Doi{10.1088/0264-9381/27/12/124005}{\bibinfo {journal} {Class. Quant
  Grav.}}\ }%
  \textbf{\bibinfo {volume} {27}},\ \bibinfo {pages} {124005} (\bibinfo {year}
  {2010}),\ \Eprint{http://arxiv.org/abs/1005.1649}{arXiv:1005.1649
  [astro-ph.CO]}%
  \bibAnnoteFile{NoStop}{Seery:2010kh}%
\bibitem{ArkaniHamed:2007ky}%
  \BibitemOpen
  \bibfield{author}{%
  \bibinfo {author} {\bibfnamefont{N.}~\bibnamefont{Arkani-Hamed}}, \bibinfo
  {author} {\bibfnamefont{S.}~\bibnamefont{Dubovsky}}, \bibinfo {author}
  {\bibfnamefont{A.}~\bibnamefont{Nicolis}}, \bibinfo {author}
  {\bibfnamefont{E.}~\bibnamefont{Trincherini}},\ and\ \bibinfo {author}
  {\bibfnamefont{G.}~\bibnamefont{Villadoro}},\ }%
  \bibfield{journal}{%
  \bibinfo {journal} {JHEP}\ }%
  \textbf{\bibinfo {volume} {05}},\ \bibinfo {pages} {055} (\bibinfo {year}
  {2007}),\ \Eprint{http://arxiv.org/abs/0704.1814}{arXiv:0704.1814 [hep-th]}%
  \bibAnnoteFile{NoStop}{ArkaniHamed:2007ky}%
\bibitem{Novello:2008ra}%
  \BibitemOpen
  \bibfield{author}{%
  \bibinfo {author} {\bibfnamefont{M.}~\bibnamefont{Novello}}\ and\ \bibinfo
  {author} {\bibfnamefont{S.~E.~P.}\ \bibnamefont{Bergliaffa}},\ }%
  \bibfield{journal}{%
  \Doi{10.1016/j.physrep.2008.04.006}{\bibinfo {journal} {Phys. Rept.}}\ }%
  \textbf{\bibinfo {volume} {463}},\ \bibinfo {pages} {127} (\bibinfo {year}
  {2008}),\ \Eprint{http://arxiv.org/abs/0802.1634}{arXiv:0802.1634
  [astro-ph]}%
  \bibAnnoteFile{NoStop}{Novello:2008ra}%
\bibitem{Borde:2001nh}%
  \BibitemOpen
  \bibfield{author}{%
  \bibinfo {author} {\bibfnamefont{A.}~\bibnamefont{Borde}}, \bibinfo {author}
  {\bibfnamefont{A.~H.}\ \bibnamefont{Guth}},\ and\ \bibinfo {author}
  {\bibfnamefont{A.}~\bibnamefont{Vilenkin}},\ }%
  \bibfield{journal}{%
  \Doi{10.1103/PhysRevLett.90.151301}{\bibinfo {journal} {Phys. Rev. Lett.}}\
  }%
  \textbf{\bibinfo {volume} {90}},\ \bibinfo {pages} {151301} (\bibinfo {year}
  {2003}),\ \Eprint{http://arxiv.org/abs/gr-qc/0110012}{arXiv:gr-qc/0110012}%
  \bibAnnoteFile{NoStop}{Borde:2001nh}%
\bibitem{Koivisto:2010jj}%
  \BibitemOpen
  \bibfield{author}{%
  \bibinfo {author} {\bibfnamefont{T.~S.}\ \bibnamefont{Koivisto}},\ }%
  \bibfield{journal}{%
  \Doi{10.1103/PhysRevD.82.044022}{\bibinfo {journal} {Phys. Rev.}}\ }%
  \textbf{\bibinfo {volume} {D82}},\ \bibinfo {pages} {044022} (\bibinfo {year}
  {2010}),\ \Eprint{http://arxiv.org/abs/1004.4298}{arXiv:1004.4298 [gr-qc]}%
  \bibAnnoteFile{NoStop}{Koivisto:2010jj}%
\bibitem{Barragan:2010qb}%
  \BibitemOpen
  \bibfield{author}{%
  \bibinfo {author} {\bibfnamefont{C.}~\bibnamefont{Barragan}}\ and\ \bibinfo
  {author} {\bibfnamefont{G.~J.}\ \bibnamefont{Olmo}}}%
   (\bibinfo {year} {2010}),\
  \Eprint{http://arxiv.org/abs/1005.4136}{arXiv:1005.4136 [gr-qc]}%
  \bibAnnoteFile{NoStop}{Barragan:2010qb}%
\bibitem{Calcagni:2010ab}%
  \BibitemOpen
  \bibfield{author}{%
  \bibinfo {author} {\bibfnamefont{G.}~\bibnamefont{Calcagni}}\ and\ \bibinfo
  {author} {\bibfnamefont{G.}~\bibnamefont{Nardelli}}}%
   (\bibinfo {year} {2010}),\
  \Eprint{http://arxiv.org/abs/1004.5144}{arXiv:1004.5144 [hep-th]}%
  \bibAnnoteFile{NoStop}{Calcagni:2010ab}%
\bibitem{Biswas:2010zk}%
  \BibitemOpen
  \bibfield{author}{%
  \bibinfo {author} {\bibfnamefont{T.}~\bibnamefont{Biswas}}, \bibinfo {author}
  {\bibfnamefont{T.}~\bibnamefont{Koivisto}},\ and\ \bibinfo {author}
  {\bibfnamefont{A.}~\bibnamefont{Mazumdar}}}%
   (\bibinfo {year} {2010}),\
  \Eprint{http://arxiv.org/abs/1005.0590}{arXiv:1005.0590 [hep-th]}%
  \bibAnnoteFile{NoStop}{Biswas:2010zk}%
\bibitem{Biswas:2010si}%
  \BibitemOpen
  \bibfield{author}{%
  \bibinfo {author} {\bibfnamefont{T.}~\bibnamefont{Biswas}}, \bibinfo {author}
  {\bibfnamefont{A.}~\bibnamefont{Mazumdar}},\ and\ \bibinfo {author}
  {\bibfnamefont{A.}~\bibnamefont{Shafieloo}}}%
   (\bibinfo {year} {2010}),\
  \Eprint{http://arxiv.org/abs/1003.3206}{arXiv:1003.3206 [hep-th]}%
  \bibAnnoteFile{NoStop}{Biswas:2010si}%
\bibitem{Biswas:2009fv}%
  \BibitemOpen
  \bibfield{author}{%
  \bibinfo {author} {\bibfnamefont{T.}~\bibnamefont{Biswas}}\ and\ \bibinfo
  {author} {\bibfnamefont{A.}~\bibnamefont{Mazumdar}},\ }%
  \bibfield{journal}{%
  \Doi{10.1103/PhysRevD.80.023519}{\bibinfo {journal} {Phys. Rev.}}\ }%
  \textbf{\bibinfo {volume} {D80}},\ \bibinfo {pages} {023519} (\bibinfo {year}
  {2009}),\ \Eprint{http://arxiv.org/abs/0901.4930}{arXiv:0901.4930 [hep-th]}%
  \bibAnnoteFile{NoStop}{Biswas:2009fv}%
\bibitem{gr}%
  \BibitemOpen
  \bibfield{author}{%
  \bibinfo {author} {\bibfnamefont{I.~S.}\ \bibnamefont{Gradshteyn}}\ and\
  \bibinfo {author} {\bibfnamefont{I.~M.}\ \bibnamefont{Ryzhik}},\ }%
  \emph{\bibinfo {title} {{Table of Integrals, Series and Products}}}\
  (\bibinfo {publisher} {New York, USA: Academic Press},\ \bibinfo {year}
  {1965})%
  \bibAnnoteFile{NoStop}{gr}%
\bibitem{Miao:2010vs}%
  \BibitemOpen
  \bibfield{author}{%
  \bibinfo {author} {\bibfnamefont{S.~P.}\ \bibnamefont{Miao}}, \bibinfo
  {author} {\bibfnamefont{N.~C.}\ \bibnamefont{Tsamis}},\ and\ \bibinfo
  {author} {\bibfnamefont{R.~P.}\ \bibnamefont{Woodard}}}%
   (\bibinfo {year} {2010}),\
  \Eprint{http://arxiv.org/abs/1002.4037}{arXiv:1002.4037 [gr-qc]}%
  \bibAnnoteFile{NoStop}{Miao:2010vs}%
\bibitem{Onemli:2002hr}%
  \BibitemOpen
  \bibfield{author}{%
  \bibinfo {author} {\bibfnamefont{V.~K.}\ \bibnamefont{Onemli}}\ and\ \bibinfo
  {author} {\bibfnamefont{R.~P.}\ \bibnamefont{Woodard}},\ }%
  \bibfield{journal}{%
  \Doi{10.1088/0264-9381/19/17/311}{\bibinfo {journal} {Class. Quant. Grav.}}\
  }%
  \textbf{\bibinfo {volume} {19}},\ \bibinfo {pages} {4607} (\bibinfo {year}
  {2002}),\ \Eprint{http://arxiv.org/abs/gr-qc/0204065}{arXiv:gr-qc/0204065}%
  \bibAnnoteFile{NoStop}{Onemli:2002hr}%
\bibitem{Kahya:2009sz}%
  \BibitemOpen
  \bibfield{author}{%
  \bibinfo {author} {\bibfnamefont{E.~O.}\ \bibnamefont{Kahya}}, \bibinfo
  {author} {\bibfnamefont{V.~K.}\ \bibnamefont{Onemli}},\ and\ \bibinfo
  {author} {\bibfnamefont{R.~P.}\ \bibnamefont{Woodard}},\ }%
  \bibfield{journal}{%
  \Doi{10.1103/PhysRevD.81.023508}{\bibinfo {journal} {Phys. Rev.}}\ }%
  \textbf{\bibinfo {volume} {D81}},\ \bibinfo {pages} {023508} (\bibinfo {year}
  {2010}),\ \Eprint{http://arxiv.org/abs/0904.4811}{arXiv:0904.4811 [gr-qc]}%
  \bibAnnoteFile{NoStop}{Kahya:2009sz}%
\bibitem{Carneiro:2001fz}%
  \BibitemOpen
  \bibfield{author}{%
  \bibinfo {author} {\bibfnamefont{S.}~\bibnamefont{Carneiro}}\ and\ \bibinfo
  {author} {\bibfnamefont{G.~A.}\ \bibnamefont{Mena~Marugan}},\ }%
  \bibfield{journal}{%
  \Doi{10.1103/PhysRevD.64.083502}{\bibinfo {journal} {Phys. Rev.}}\ }%
  \textbf{\bibinfo {volume} {D64}},\ \bibinfo {pages} {083502} (\bibinfo {year}
  {2001}),\ \Eprint{http://arxiv.org/abs/gr-qc/0109039}{arXiv:gr-qc/0109039}%
  \bibAnnoteFile{NoStop}{Carneiro:2001fz}%
\bibitem{Koivisto:2010dr}%
  \BibitemOpen
  \bibfield{author}{%
  \bibinfo {author} {\bibfnamefont{T.~S.}\ \bibnamefont{Koivisto}}, \bibinfo
  {author} {\bibfnamefont{D.~F.}\ \bibnamefont{Mota}}, \bibinfo {author}
  {\bibfnamefont{M.}~\bibnamefont{Quartin}},\ and\ \bibinfo {author}
  {\bibfnamefont{T.~G.}\ \bibnamefont{Zlosnik}}}%
   (\bibinfo {year} {2010}),\
  \Eprint{http://arxiv.org/abs/1006.3321}{arXiv:1006.3321 [astro-ph.CO]}%
  \bibAnnoteFile{NoStop}{Koivisto:2010dr}%
\bibitem{Nojiri:2005sx}%
  \BibitemOpen
  \bibfield{author}{%
  \bibinfo {author} {\bibfnamefont{S.}~\bibnamefont{Nojiri}}, \bibinfo {author}
  {\bibfnamefont{S.~D.}\ \bibnamefont{Odintsov}},\ and\ \bibinfo {author}
  {\bibfnamefont{S.}~\bibnamefont{Tsujikawa}},\ }%
  \bibfield{journal}{%
  \Doi{10.1103/PhysRevD.71.063004}{\bibinfo {journal} {Phys. Rev.}}\ }%
  \textbf{\bibinfo {volume} {D71}},\ \bibinfo {pages} {063004} (\bibinfo {year}
  {2005}),\ \Eprint{http://arxiv.org/abs/hep-th/0501025}{arXiv:hep-th/0501025}%
  \bibAnnoteFile{NoStop}{Nojiri:2005sx}%
\bibitem{Tsamis:1996qq}%
  \BibitemOpen
  \bibfield{author}{%
  \bibinfo {author} {\bibfnamefont{N.~C.}\ \bibnamefont{Tsamis}}\ and\ \bibinfo
  {author} {\bibfnamefont{R.~P.}\ \bibnamefont{Woodard}},\ }%
  \bibfield{journal}{%
  \Doi{10.1016/0550-3213(96)00246-5}{\bibinfo {journal} {Nucl. Phys.}}\ }%
  \textbf{\bibinfo {volume} {B474}},\ \bibinfo {pages} {235} (\bibinfo {year}
  {1996}),\ \Eprint{http://arxiv.org/abs/hep-ph/9602315}{arXiv:hep-ph/9602315}%
  \bibAnnoteFile{NoStop}{Tsamis:1996qq}%
\bibitem{Abramo:1997hu}%
  \BibitemOpen
  \bibfield{author}{%
  \bibinfo {author} {\bibfnamefont{L.~R.~W.}\ \bibnamefont{Abramo}}, \bibinfo
  {author} {\bibfnamefont{R.~H.}\ \bibnamefont{Brandenberger}},\ and\ \bibinfo
  {author} {\bibfnamefont{V.~F.}\ \bibnamefont{Mukhanov}},\ }%
  \bibfield{journal}{%
  \Doi{10.1103/PhysRevD.56.3248}{\bibinfo {journal} {Phys. Rev.}}\ }%
  \textbf{\bibinfo {volume} {D56}},\ \bibinfo {pages} {3248} (\bibinfo {year}
  {1997}),\ \Eprint{http://arxiv.org/abs/gr-qc/9704037}{arXiv:gr-qc/9704037}%
  \bibAnnoteFile{NoStop}{Abramo:1997hu}%
\bibitem{Harrison:1967zz}%
  \BibitemOpen
  \bibfield{author}{%
  \bibinfo {author} {\bibfnamefont{E.~R.}\ \bibnamefont{Harrison}},\ }%
  \bibfield{journal}{%
  \Doi{10.1103/RevModPhys.39.862}{\bibinfo {journal} {Rev. Mod. Phys.}}\ }%
  \textbf{\bibinfo {volume} {39}},\ \bibinfo {pages} {862} (\bibinfo {year}
  {1967})%
  \bibAnnoteFile{NoStop}{Harrison:1967zz}%
\bibitem{Lyth:1995cw}%
  \BibitemOpen
  \bibfield{author}{%
  \bibinfo {author} {\bibfnamefont{D.~H.}\ \bibnamefont{Lyth}}\ and\ \bibinfo
  {author} {\bibfnamefont{A.}~\bibnamefont{Woszczyna}},\ }%
  \bibfield{journal}{%
  \Doi{10.1103/PhysRevD.52.3338}{\bibinfo {journal} {Phys. Rev.}}\ }%
  \textbf{\bibinfo {volume} {D52}},\ \bibinfo {pages} {3338} (\bibinfo {year}
  {1995}),\
  \Eprint{http://arxiv.org/abs/astro-ph/9501044}{arXiv:astro-ph/9501044}%
  \bibAnnoteFile{NoStop}{Lyth:1995cw}%
\end{thebibliography}%

\end{document}